\documentclass[12pt]{article}
\usepackage{epsfig, epsf, graphicx, subfigure, color}
\usepackage{pstricks, pst-node, psfrag}
\usepackage{amssymb,amsmath}
\usepackage{verbatim,enumerate}
\usepackage{rotating, lscape}
\usepackage{setspace}
\usepackage{bbm}
\usepackage{natbib}
\usepackage{amsthm}
\usepackage{bm}
\usepackage{url}
\usepackage{appendix}
\usepackage{multirow}
\usepackage{xspace}

\usepackage{tabularx}
\newcolumntype{Y}{>{\centering\arraybackslash}X} 

\usepackage{authblk} 

\theoremstyle{definition}

\newtheorem{Remark}{Remark}

\newcommand{\plasma}{\texttt{PLASMA}\xspace}
\newcommand{\blas}{\texttt{BLAS}\xspace}

\newcommand{\starpu}{\texttt{StarPU}\xspace}

\newcommand{\chameleon}{\texttt{Chameleon}\xspace}
\newcommand{\exageostat}{\texttt{ExaGeoStat}\xspace}
\newcommand{\exageostatr}{\texttt{ExaGeoStatR}\xspace}
\newcommand{\hicma}{\texttt{HiCMA}\xspace}

\newcommand\citepa[1]{(\citeauthor{#1},~2019a)}
\newcommand\citec[1]{\citeauthor{#1}~(2019c)}
\newcommand\citepb[1]{(\citeauthor{#1},~2019b)}

\newcommand{\bi}{\begin{itemize}}
	\newcommand{\ei}{\end{itemize}}

\usepackage{lineno}

\begin{document}
	

\graphicspath{{figures/}}	
	
	\thispagestyle{empty} \baselineskip=28pt \vskip 5mm
	\begin{center} {\Large{\bf Efficiency Assessment of Approximated Spatial Predictions for Large Datasets}}
	\end{center}
	
	\baselineskip=12pt \vskip 5mm








	\begin{center}\large
		Yiping Hong\footnote{\baselineskip=10pt 
			Statistics Program, King Abdullah University of Science and Technology, Thuwal 23955-6900, Saudi Arabia. E-mails: yiping.hong@kaust.edu.sa (Yiping Hong),  marc.genton@kaust.edu.sa (Marc G. Genton), ying.sun@kaust.edu.sa (Ying Sun). 
		}, Sameh Abdulah\footnote[2]{\baselineskip=10pt 
		Extreme Computing Research Center (ECRC), King Abdullah University of Science and Technology, Thuwal 23955-6900, Saudi Arabia. 
		E-mail: sameh.abdulah@kaust.edu.sa. 
	    }, 
	    Marc G. Genton\footnotemark[\value{footnote}], 
		and Ying Sun\footnotemark[\value{footnote}]
		\end{center}
	
	\baselineskip=16pt \vskip 1mm \centerline{\today} \vskip 8mm
	
		{\large{\bf Abstract: }}
Due to the well-known computational showstopper of the exact Maximum Likelihood Estimation (MLE) for large geospatial observations, 
a variety of approximation methods have been proposed in the literature, 
which usually require tuning certain inputs. 
For example, the recently developed Tile Low-Rank approximation (TLR) method involves many tuning parameters, including numerical accuracy. 
To properly choose the tuning parameters, 
it is crucial to adopt a meaningful criterion for the assessment of the prediction efficiency with different inputs. 
Unfortunately, the most commonly-used Mean Square Prediction Error (MSPE) criterion cannot directly assess the loss of efficiency when the spatial covariance model is approximated. 
Though the Kullback-Leibler Divergence criterion can provide the information loss of the approximated model, it cannot give more detailed information that one may be  interested in, e.g., the accuracy of the computed MSE. 
In this paper, we present three other criteria, the
Mean Loss of Efficiency (MLOE), Mean Misspecification of the Mean Square Error (MMOM), and Root mean square MOM (RMOM), 
and show numerically that, in comparison with the common MSPE criterion and the Kullback-Leibler Divergence criterion, our criteria are more informative, and thus more 
adequate to
assess the loss of the prediction efficiency by using the approximated or misspecified covariance models.
Hence, our suggested criteria are more useful for the 
determination of tuning parameters for sophisticated approximation methods of spatial model fitting. 
To illustrate this, we investigate the trade-off between the execution time, estimation
accuracy, and prediction efficiency for the TLR method with extensive simulation studies
and suggest proper settings of the TLR tuning parameters. 
We then apply the TLR method to a large spatial dataset of soil moisture in the area of the Mississippi River basin, 
and compare the TLR with the Gaussian predictive process and the composite likelihood method, showing that our suggested criteria can successfully be used to choose the tuning parameters that can keep the estimation or the prediction accuracy in applications. 
		
	\baselineskip=17pt

	\begin{doublespace}
		
		\par\vfill\noindent
	\par\medskip\noindent
	\end{doublespace}
		

\baselineskip=25pt 

\setlength\abovedisplayskip{1pt}
\setlength\belowdisplayskip{1pt}

\section{Introduction}
\label{sec:intro}

Geostatistical applications include modeling the spatial distribution of a set
of observations (e.g., temperature, humidity, soil moisture, wind speed) 
taken at $n$ locations regularly or irregularly spaced over a 
given geographical area. 
In geostatistics, the spatial datasets are often considered as a realization of a Gaussian process, 
defined by a mean function and a spatial covariance model. 
More specifically, we suppose that the data are observed from a stationary, isotropic Gaussian random field $\{ Z(\bm{s}): \bm{s}\in D \subset \mathbb{R}^d \}$, 
with mean zero and covariance function $C(h; \bm{\theta}) := \text{Cov}_{\bm{\theta}}\{ Z(\bm{s}_1) , Z(\bm{s}_2) \}$ for any $\bm{s}_1, \bm{s}_2 \in D$ and $\| \bm{s}_1 - \bm{s}_2 \| = h$, where $\bm{\theta}$ is the unknown parameter vector. 
In recent years, the Mat\'{e}rn family has been a popular choice for the covariance function, 
since it represents
a general form of many possible covariance models in the literature, due to its flexibility.
The Mat\'{e}rn covariance function is defined as
\begin{equation} \label{Eq:Matern_Cov}
C(h; \bm{\theta}) = \frac{\sigma^2}{\Gamma(\nu) 2^{\nu - 1} } \left(\frac{h}{\alpha}\right)^\nu \mathcal{K}_\nu \left(  \frac{h}{\alpha} \right), 
\end{equation} 
where $\bm{\theta} = (\sigma^2, \alpha, \nu)^\top$, $\sigma^2>0$, $\alpha>0$, and $\nu>0$ are the variance, range parameter, and smoothness parameter, respectively, 
and $\mathcal{K}_\nu$ is the modified Bessel function of the second kind of order $\nu$.

The Maximum Likelihood Estimation (MLE) method
has been widely used for estimating the parameter vector $\bm{\theta}$ of the spatial model.
Denoting the spatial dataset by $\bm{Z} = \{ Z(\bm{s}_1), \ldots, Z(\bm{s}_n) \}^\top$, 
where $\bm{s}_1, \ldots, \bm{s}_n$ are the observation locations, 
the MLE of the unknown parameter $\bm{\theta}$ can then be obtained by maximizing the following log-likelihood function: 
\begin{equation} \label{Eq:Log_likelihood}
l(\bm{\theta}) = -\frac{n}{2} \log(2\pi) - \frac{1}{2} \log \det \{ \Sigma(\bm{\theta}) \} - \frac{1}{2} \bm{Z}^\top \Sigma(\bm{\theta})^{-1} \bm{Z}, 
\end{equation}
where $\Sigma(\bm{\theta})$ is the covariance matrix, with entries 
$\left[ \Sigma(\bm{\theta}) \right]_{i, j} = C(\| \bm{s}_i-\bm{s}_j \|; \bm{\theta})$ for $i, j = 1, \ldots, n$. 
Finding the exact MLE requires $O(n^3)$ computations and $O(n^2)$ memory, 
since evaluating the log-likelihood function involves the inverse and the determinant of the covariance matrix. 
Thus, the exact MLE is not feasible for large spatial datasets in applications,
e.g. meteorological data, where $n$ is often of an order of $10^5$ or $10^6$.

To overcome this computational problem, finding approximation methods to compute the
MLE has drawn considerable attention. 
The approximation can be applied to the spatial model, log-likelihood function, and covariance matrix. 
First, the spatial model can be approximated by a low-rank model, which is easier to compute. For instance, \cite{cressie2006spatial, cressie2008fixed} proposed the fixed rank kriging (FRK) method, which approximates the spatial dependence model by a linear combination of proper basis functions. 
\cite{banerjee2008gaussian} introduced the Gaussian predictive process (GPP), 
where the spatial model is approximated by the kriging prediction using the observations on some pre-determined knots plus a nugget effect. \cite{finley2009improving} modified this method by introducing the fine-scale process and fixed the problem that the marginal variance is underestimated. 
Second, for the approximation of log-likelihood function, \cite{vecchia1988estimation} and \cite{curriero1999composite} introduced the composite likelihood approach by ignoring the correlation of the observations at distant locations in the function. 
\cite{stein2004approximating} showed that this approximation could also be adapted to the restricted likelihood. 

Third, the covariance matrix can be approximated by a sparse matrix. 
In the covariance tapering method \citep{furrer2006covariance, kaufman2008covariance, du2009fixed}, the covariance matrix is multiplied element-wise by a sparse covariance matrix, so the dependency between distant locations are neglected. 
\cite{stein2014limitations} showed that one could approximate the covariance matrix by dividing the covariance matrix by several tiles and replacing the off-diagonal tiles by zero matrices. This approximation can provide a more accurate prediction compared with the low-rank model-based method. Naturally, one can introduce a more delicate sparse structure for covariance matrix approximation. 
The H-matrix \citep{hackbusch1999sparse} defines a hierarchical block structure for the matrix, which allows a coarse approximation for the block distant from the diagonal and a delicate approximation for the block near the diagonal. There are different kinds of H-matrix approximation, such as the HODLR \citep{aminfar2016fast}, HSS \citep{ghysels2016efficient} , H2-matrices \citep{borm2016approximation, sushnikova2016preconditioners}, and BLR/TLR \citep{pichon2017sparse, akbudak2017tile, abdulah2018tile}. 
The recently proposed Tile Low-Rank (TLR) approximation method \citep{akbudak2017tile, abdulah2018tile} divides the covariance matrix into several tiles and performs low-rank approximations
on the off-diagonal tiles. \cite{abdulah2018tile} showed that it could improve the computation of the likelihood function
on parallel architectures such as shared-memory, GPUs, and distributed-memory systems. \citec{abdulah2019geostatistical} also considered using different precisions for the diagonal and off-diagonal tiles in the Cholesky decomposition of covariance matrices, which can also improve the computational performance. 
One can also approximate the inverse of the covariance matrix, or the precision matrix, instead \citep{lindgren2011explicit, nychka2015multiresolution}.  \cite{sun2016statistically} introduced a sparse inverse Cholesky decomposition in the score equation and obtained the score equation approximation method. 
Besides the categories stated above, the MLE can also be approximated by algorithmic approaches, such as the metakriging \citep{minsker2014robust}, the gapfill method \citep{gerber2018predicting}, and the local approximate Gaussian process \citep{gramacy2015local}. 
For a detailed review of the MLE approximation approaches in the literature, refer to \cite{sun2012geostatistics} and \cite{heaton2019case}.

All the above approximation methods require certain types of tuning, to some extent. 
We can call them `tuning parameters' to distinguish them from model parameters that need to be estimated from the data.
For instance, for the covariance tapering method \citep{kaufman2008covariance}, 
the taper range is a tuning parameter. 
The composite likelihood method \citep{vecchia1988estimation} approximates the conditional density $p( \bm{s}_i|\bm{s}_1, \ldots, \bm{s}_{i-1} )$  conditioning on a subset of $\bm{s}_1,\ldots,\bm{s}_{i-1}$, such as $m$ nearest neighbors of $\bm{s}_i$, for which $m$ is a tuning parameter.
In the Gaussian predictive process model \citep{banerjee2008gaussian}, the predetermined knots are tuning parameters.
The TLR approximation \citep{abdulah2018tile} involves many tuning parameters, such as the matrix tile size, TLR maximum rank, TLR numerical accuracy, and optimization tolerance, which are introduced in Section \ref{sec:background}. 
The tuning parameters should balance the computational burden and the estimation or prediction accuracy, 
so it is crucial to understand the impact of the tuning parameters on the statistical properties of the approximation methods. 
Finding a suggestion for these parameters serves as a motivation of our research, i.e., we would like to tune the TLR method input parameters, which can cut the computational time without losing too much estimation or prediction performance. 
The estimation performance is often evaluated via summary statistics and the plot of the estimations, 
such as the estimation variance \citep{kaufman2008covariance} 
or the boxplots \citep{abdulah2018tile}, 
but the prediction performance is not so straightforward to assess.

In the literature, the prediction performance is often evaluated by cross-validation. 
This method randomly leaves out $p$ locations $\bm{s}_1, \ldots, \bm{s}_p$ from observation locations, and predicts the $Z(\bm{s}_1), \ldots, Z(\bm{s}_p)$ using the rest of the data at all other locations. 
Denote these predictions by $\hat{Z}(\bm{s}_1), \ldots, \hat{Z}(\bm{s}_p)$. The prediction performance is assessed by the deviation between the true and the predicted values, such as the Mean Square Prediction/Kriging Error (MSPE) \citep{abdulah2018tile}
\begin{equation} \label{Eq:Def_MSPE}
\text{MSPE} = \frac{1}{p} \sum_{i=1}^p  \{ \hat{Z}(\bm{s}_i) - Z(\bm{s}_i) \}^2, 
\end{equation} 
or the Mean Square Relative Prediction Error \citep{yan2018gaussian} 
\begin{equation*}
\text{MSRPE} = \frac{1}{p} \sum_{i=1}^p \left\{ \frac{\hat{Z}(\bm{s}_i) - Z(\bm{s}_i)}{Z(\bm{s}_i)} \right\}^2. 
\end{equation*}
This performance can also be assessed by the deviation between the true observations and the corresponding predicted distributions, such as various kinds of proper scoring rules defined by \cite{gneiting2007strictly}. 
For prediction intervals, the performance can be assessed by 
the empirical coverage of 95\% prediction intervals on the left-out locations \citep{banerjee2008gaussian} or the interval score \citep{gneiting2007strictly}. 
For more cross-validation based criteria, see \cite{dai2007spatial}, \cite{hengl2004generic}, and \cite{heaton2019case}. 
These criteria provide a straightforward measure of the performance of the prediction. 
However, they do not directly assess the loss of statistical efficiency when the approximated model
is adopted instead of the true model, 
such as the extra Mean Square Errors (MSEs) caused by using the approximated model and the accuracy of estimated MSEs.

In the context of covariance model misspecification, \cite{stein1999interpolation} 
proposed the Loss of Efficiency (LOE) and the Misspecification of the MSE
(MOM) criteria, based on the comparison of the MSEs
between the true and the misspecified models. Using these criteria,  \cite{stein1999interpolation} deduced that the simple kriging prediction
is asymptotically optimal when the misspecified covariance model is equivalent
to the true model. 
\cite{stein1999interpolation} also performed some simulations to assess the prediction performance of the kriging prediction under different
settings of observation locations. 
However, all the results presented in \cite{stein1999interpolation} are for the case of a single prediction location.

In this article, we aim to give more appropriate criteria for the assessment of the loss of prediction efficiency when the true covariance model is approximated. 
Our suggested criteria can be used to assess the prediction efficiency of the approximation methods, e.g., the TLR method, with different tuning parameters, and help to choose the best value of these parameters. 
We suggest using the Mean Loss of Efficiency (MLOE) and 
the Mean Misspecification of the Mean Square Error (MMOM) criteria for multiple prediction locations 
as a generalization of the criteria proposed by \cite{stein1999interpolation}. 
Here the MLOE and MMOM are relative errors. MLOE is strictly positive, while MMOM can be positive or negative at different locations. 
To avoid the possible issue of the cancellation of error over multiple locations in the MMOM criterion, we also introduce the Root mean square MOM (RMOM) criterion to evaluate the deviance of MOM from zero. 
Since the approximated covariance model can be viewed as a type of model misspecification, 
to show the MLOE, MMOM, and RMOM criteria are appropriate to assess the loss of prediction efficiency, 
we perform a similar simulation study from \cite{stein1999interpolation}, where 
the exponential covariance model is misspecified as a Whittle covariance
model plus a nugget effect, 
implying that the approximated covariance is smoother than the truth.
Numerical results show that our criteria are better in assessing the prediction efficiency 
than the commonly used MSPE criterion and the Kullback-Leibler divergence criterion, which can be deduced from the logarithm score in \cite{gneiting2007strictly}. 
As an application of our suggested criteria and a response to our research motivation, 
we use them to give a practical suggestion for selecting the tuning parameters in the TLR method,  for which we investigate 
the performance of prediction and computation, using different
tuning parameters from extensive simulation studies. 
For illustration of the validity of our suggested TLR tuning parameters, 
we fit a Gaussian-process model with a Mat\'{e}rn covariance function to a large spatial dataset of soil moisture in the area of the Mississippi basin; 
we then apply the TLR approximation method to obtain the MLEs and perform predictions with the suggested tuning parameters. 
Results show that our criteria are capable of selecting the tuning parameters of the TLR approximation since the TLR works well with our suggested parameters for the soil dataset. We also compare the TLR with the composite likelihood \citep{vecchia1988estimation} and the Gaussian predictive process \citep{banerjee2008gaussian} for reference, suggesting that our criteria are successfully applied to the TLR tuning parameter selection in application. 
To the best of our knowledge, there is no previous work that considered choosing tuning parameters of an approximation method using a simulation-based criterion, rather than a data-based criterion such as the MSPE, for prediction performances.

The original LOE and MOM criteria proposed by \cite{stein1999interpolation} are for the assessment of model misspecification. We obtain its mean version for the understanding of the tuning parameters in approximation methods for large spatial datasets, which can be used to help determine the tuning parameters in different approximation methods for real applications. For instance, to determine the tapering range in the covariance tapering method, one can first 
run a simulation for moderate datasets, using our criteria to compare the prediction efficiency with different tapering ranges. The best tapering range may be dependent on the range parameter $\alpha$. 
Once the simulation results are obtained, one can choose the tapering range in the approximation problem for different real datasets, based on the simulation and a rough estimate of $\alpha$. 
However, more advanced and accurate methods often involve more tuning parameters that require intensive simulation studies to understand the impact of each parameter and determine the tuning parameters that can provide the best trade-off between statistical properties and computational cost. Therefore we use a more delicate method (TLR) to show the effectiveness of our criteria. 

The remainder of this article is organized as follows: 
Section \ref{sec:background} gives a brief background on the TLR approximation method and its tuning parameters. 
Section \ref{sec:method} introduces our suggested MLOE, MMOM, and RMOM criteria. 
In Section \ref{Sec:Simulation_Misspecified}, we perform a simulation of the validity and sensitivity of the suggested criteria. 
In Section \ref{sec:experimental_results}, we explain the simulation design to assess the TLR method, using different tuning parameters settings for which, 
our suggested criteria are used to measure the prediction accuracy and select the best specification of those tuning parameters. 
Section \ref{sec:application} shows the effectiveness of our suggested tuning parameters for the TLR method, using the real soil moisture dataset. For this dataset, we also compare the estimation and prediction efficiency for the TLR method with our suggested parameters, with the composite likelihood and the Gaussian predictive process method in this section. 
Conclusions and discussions are provided in Section \ref{sec:discussion}. 
More detailed numerical results about the specification of tuning parameters for the TLR method 
can be found in the Supplementary Material. 
\section{Tile Low-Rank (TLR) Approximation}
\label{sec:background}
In this section, we give a brief background on the TLR approximation method, together with the tuning parameters associated with it in  parallel hardware environments. 

Tile-based algorithms have been developed on parallel architectures to speedup matrix-linear solver algorithms,
for instance, \plasma ~\citep{agullo2009numerical} and \chameleon ~\citep{chameleon} libraries.
The given matrix is split into a set of tiles to allow the use of parallel execution, to a maximum degree, 
by weakening the synchronization points and bringing the parallelism in multithreaded \blas \citep{blas2002} to maximize the hardware utilization.

Since maximizing the log-likelihood in \eqref{Eq:Log_likelihood} and obtaining the MLE involves applying a set of linear-solver operations to the geospatial covariance matrix $\Sigma$, 
\cite{abdulah2018exageostat} have developed \exageostat ~\footnote{https://github.com/ecrc/exageostat}, 
a framework that use tile-based linear algebra algorithms to parallelize the MLE operations on leading-edge parallel hardware architecture.  This framework has also been extended in~\cite{abdulah2018tile} to apply
a TLR approximation to the covariance matrix. The new approximation technique
aims at exploiting the data sparsity of the dense covariance matrix by compressing the off-diagonal tiles up to a user-defined
accuracy threshold. The TLR method differs from  existing low-rank approximation techniques, e.g. \cite{banerjee2008gaussian},  as the low-rank approximation is applied separately
on each tile, instead of the whole matrix.

\begin{figure}[htbp]
    \centering
    \includegraphics[width=0.75\textwidth]{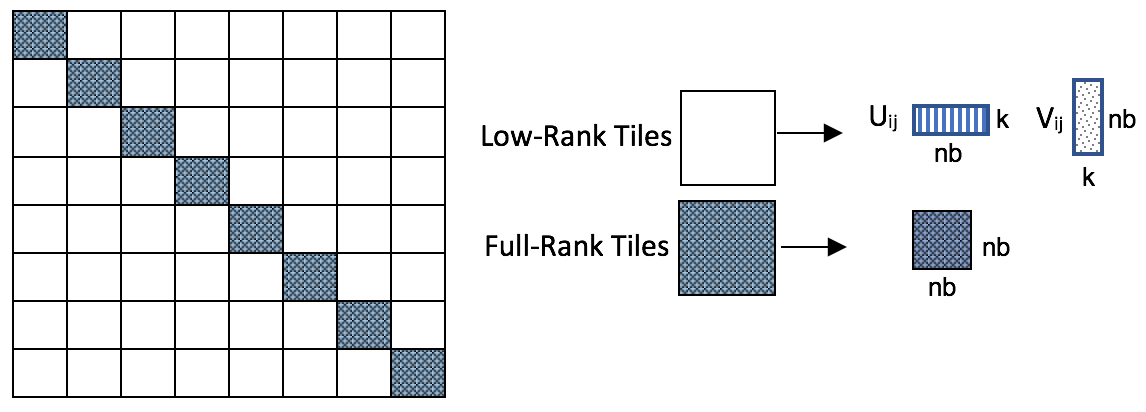}
    \caption{An illustrative example of a 8 $\times$ 8 covariance matrix TLR structure.}
    \label{Fig:TLR_Idea}
\end{figure}

Figure \ref{Fig:TLR_Idea} gives an illustrative example of the TLR approximation method to a $8 \times 8$ covariance
matrix, e.g., $\Sigma(\bm{\theta})$, where $\bm{\theta}$ represents the parameter vector (i.e., variance, range,
and smoothness parameters in the Mat\'{e}rn covariance function). 
Assuming a square positive-definite covariance matrix, the spatial covariance matrix with size $n \times n$ is divided
into several tiles $D_{i, j}(\bm{\theta})$, where the size of each tile is $nb \times nb$. 
The Singular Value Decomposition (SVD) is used to approximate the off-diagonal tiles to a user-defined
accuracy (i.e., $tlr\_acc$, the tuning parameter argument used in \exageostat as indicated in Table \ref{table:nonlin}). 
In this case, the approximated tiles are the multiplication of two low-rank
matrices, e.g., $D_{i, j}(\bm{\theta})$ is approximated by $\tilde{D}_{i, j}(\bm{\theta})=U_{i, j}(\bm{\theta})V_{i, j}(\bm{\theta})$, which
can be deduced from the $k$ most significant singular values and their associated left and right singular vectors.

This approximation gives a data compression format that requires less memory and offers a faster
computational speed of the matrix algebra. 
In the \exageostat software \citep{abdulah2018tile}, the TLR approximation is performed
by the Hierarchical Computations on Manycore Architectures (\hicma)
numerical library~\citepa{hicma2019}, which allows to run the approximation on parallel systems with the help of \starpu \citep{augonnet2011starpu}.

Applying the TLR approximation to the log-likelihood function requires tuning several inputs to control the performance and accuracy
of the approximation, namely, $nb$, $tlr\_max\_rank$, $tlr\_acc$, and $opt\_tol$ as shown in Table \ref{table:nonlin}; $nb$
controls the size of each tile $D_{i, j}(\bm{\theta})$, and $tlr\_max\_rank$ determines the maximum possible rank of the approximated tiles,
which affects the memory allocation process of the approximating low-rank matrices $U_{i, j}(\bm{\theta})$ and $V_{i, j}(\bm{\theta})$ in the \hicma library. 
By adopting the suggested criteria when assessing the prediction efficiency, 
we herein determine the best combination of the four TLR inputs by tuning these inputs, and by evaluating the
performance and the accuracy of the approximated MLE compared with the exact MLE solution.

\tabcolsep=0.11cm
\begin{table}[ht]
	\small
	\caption{Arguments for the tuning parameters of the TLR method in the \exageostat framework} 
	\centering 
	\begin{tabular}{|c|c| } 
		\hline\hline 
		Name & Symbol \\ [1 ex] 
		\hline 
		Matrix tile size & $nb$  \\ 
		TLR maximum rank & $tlr\_max\_rank$\\
		TLR numerical accuracy & $tlr\_acc$\\
		Optimization tolerance & $opt\_tol$\\

		\hline 
	\end{tabular}
	\label{table:nonlin} 
\end{table}

The effectiveness of the TLR approximation method can be improved by well tuning these four inputs. For instance, the current
implementation of TLR in \hicma uses a fixed-rank method to allocate and process all the given matrix tiles, although different
approximated tiles have different ranks. A value of $tlr\_max\_rank$ that is too large causes unnecessary memory usage and more
data movements in the case of distributed memory architectures, whereas a too small value may cause a failure in approximating the tile.
Thus, the best value of $tlr\_max\_rank$ should be the smallest possible value that makes the approximation feasible for all the off-diagonal tiles. 
The accuracy threshold $tlr\_acc$
is also important to control the approximation accuracy, such that the approximation $\tilde{D}_{i, j}(\bm{\theta})$ of each tile
satisfies $\| \tilde{D}_{i, j}(\bm{\theta})-D_{i, j}(\bm{\theta})\|_2 \le tlr\_acc$, where $\|\cdot\|_2$ is the $L_2$-norm of a matrix. A lower accuracy
(larger $tlr\_acc$) brings the arithmetic intensity of the approximation close to the memory-bound regime, whereas a higher accuracy
makes the approximation run in the compute-bound regime \citep{abdulah2018tile}. Thus, the accuracy threshold is an application-specific value. 
Furthermore, the optimization tolerance $opt\_tol$ is the minimum difference between two log-likelihood values at different iterations
to control the optimization convergence condition. More specifically, the iteration process
of computing the maximum point stops when $|l(\bm{\theta}_{\text{opt}})-l(\bm{\theta}_{\text{sub}})|\le opt\_tol$, where $l(\bm{\theta}_{\text{opt}})$ is the largest value
of the log-likelihood function over all iterations and $l(\bm{\theta}_{\text{sub}})$ is the second largest one.

We hope the tuning parameters in the TLR can save as much computational time as possible,
without losing too much in estimating the accuracy of the prediction. \cite{abdulah2018tile} investigated the impact of $tlr\_acc$ by
showing the boxplots of the estimated parameters and the MSPEs. 
Here, our work uses the more informative MLOE, MMOM, and RMOM 
criteria for assessing the spatial prediction efficiency, which we describe in more details in Section \ref{sec:method}.

In this study, we use the \exageostatr~\footnote{https://github.com/ecrc/exageostatR} package to perform the 
experiments relating to the TLR approximation. \exageostatr \citepb{abdulah2019exageostatr} is the \textsf{R}-wrapper interface of \exageostat developed to facilitate the exploitation of large-scale capabilities in the \textsf{R}\xspace environment. 
The package provides parallel computation for the evaluation of the Gaussian maximum likelihood function using shared memory, GPUs, and distributed systems, by mitigating its memory space and computing restrictions. This package
provides three ways of computing the MLE on a large scale: exact, Diagonal Super Tile (DST) approximation (i.e., covariance tapering), and TLR approximation.  We are targeting the \textsf{R}\xspace functions related to the TLR approximation.
The function $\texttt{tlr\_mle()}$ in the \exageostatr package allows the computation of the TLR approximation of
the MLE for the Mat\'{e}rn covariance model. This function computes the estimation by substituting the covariance matrix with
its TLR approximation in the exact MLE framework.

\section{Efficiency Criteria for Approximated Spatial Predictions} 
\label{sec:method}

In this section, we construct three criteria for assessing the accuracy of the spatial prediction, when the covariance matrix in the log-likelihood in \eqref{Eq:Log_likelihood} is approximated.
Our first two criteria are of the averaged form of the criteria called the Loss of Efficiency (LOE) and the Misspecification of the MSE (MOM), proposed by \cite{stein1999interpolation},
in the context of spatial prediction with a misspecified covariance model. Our last criterion is of the mean square form of the MOM to measure the MOM variability on different prediction locations. 

We consider a zero-mean Gaussian random field $Z(\bm{s})$, where the observations are $\bm{Z} = \left\{ Z(\bm{s}_1), \ldots, Z(\bm{s}_n)\right\}^\top$. When the covariance model is true, the kriging prediction of $Z(\bm{s}_0)$ at a point $\bm{s}_0$ is 
$ 
\hat{Z}_t(\bm{s}_0) = \bm{k}_t^\top K_t^{-1} \bm{Z}, 
$
with MSE given by 
$ 
\text{MSE}(\bm{s}_0) = \text{E}_t \{ e_t^2(\bm{s}_0) \} = k_{0t} - \bm{k}_t^\top K_t^{-1} \bm{k}_t, 
$
where $e_t(\bm{s}_0) = \hat{Z}_t(\bm{s}_0) - Z(\bm{s}_0)$ is the error of the kriging predictor, $K_t = \text{Cov}_t \{ \bm{Z}, \bm{Z}^\top \}$, $\bm{k}_t = \text{Cov}_t \{ \bm{Z}, Z(\bm{s}_0) \}$, $k_{0t} = \text{Var}_t \{ Z(\bm{s}_0) \}$, $\text{E}_t$, $\text{Var}_t$, and $\text{Cov}_t$ mean the 
expectation, variance, and covariance with respect to the true covariance model. 
However, when the covariance is approximated, the kriging predictor is 
$\hat{Z}_a(\bm{s}_0) = \bm{k}_a^\top K_a^{-1} \bm{Z}$ 
instead, where $K_a = \text{Cov}_a \{ \bm{Z}, \bm{Z}^\top \}$, $\bm{k}_a = \text{Cov}_a \{ \bm{Z}, Z(\bm{s}_0) \}$, and $\text{Cov}_a$ means the covariance is computed under the approximated 
covariance model. Denoting the error of this predictor by $e_a(\bm{s}_0) = \hat{Z}_a(\bm{s}_0) - Z(\bm{s}_0)$, then the MSE of this prediction is actually 
$
\text{E}_t \{ e_a^2(\bm{s}_0) \} = k_{0t} - 2\bm{k}_t^\top K_a^{-1} \bm{k}_a + \bm{k}_a^\top K_a^{-1} K_t K_a^{-1} \bm{k}_a, 
$
and the calculated result of MSE is 
$
\text{E}_a \{ e_a^2(\bm{s}_0) \} = k_{0a} - \bm{k}_a^\top K_a^{-1} \bm{k}_a, 
$
where $k_{0a} = \text{Var}_a \{ Z(\bm{s}_0) \}$, $\text{E}_a$ and $\text{Var}_a$ mean that the expectation and variance are computed using the  approximated covariance model. 
Thus, following \cite{stein1999interpolation}, the Loss of Efficiency of the prediction is defined as 
\begin{equation} \label{Eq:LOE_one_point}
\text{LOE}(\bm{s}_0) = \text{E}_t \{ e_a^2(\bm{s}_0) \} / \text{E}_t \{ e_t^2(\bm{s}_0) \} - 1, 
\end{equation}
and the Misspecification of the MSE is defined as 
\begin{equation} \label{Eq:MOM_one_point}
\text{MOM}(\bm{s}_0) = \text{E}_a \{ e_a^2(\bm{s}_0) \} / \text{E}_t \{ e_a^2(\bm{s}_0) \} - 1. 
\end{equation}

Our first two criteria are defined as the mean value of the Loss of Efficiency \eqref{Eq:LOE_one_point} and Misspecification of the MSE \eqref{Eq:MOM_one_point} over multiple prediction locations. 
More specifically, when the prediction locations are $\bm{s}_{01}, \ldots, \bm{s}_{0m}$, the Mean Loss of Efficiency is defined as 
\begin{equation} \label{Eq:MLOE}
\text{MLOE} = \frac{1}{m}\sum_{i=1}^m \text{LOE}(\bm{s}_{0i}), 
\end{equation}
and the Mean Misspecification of the MSE is defined as 
\begin{equation} \label{Eq:MMOM}
\text{MMOM} = \frac{1}{m}\sum_{i=1}^m \text{MOM}(\bm{s}_{0i}). 
\end{equation}

For Gaussian random fields, the kriging predictor $\hat{Z}_t(\bm{s}_0)$ is the best predictor in terms of minimizing MSE \citep{stein1999interpolation}, so $\text{E}_t \{ e_t^2(\bm{s}_0) \} \le \text{E}_t \{ e_a^2(\bm{s}_0) \}$ and $\text{LOE}(\bm{s}_0) \ge 0$ for any $\bm{s}_0$. However, $\text{MOM}(\bm{s}_{0i})$ may be positive or negative. If there are two $\text{MOM}$ values which have opposite sign and large absolute values, they will eliminate each other in \eqref{Eq:MMOM}, causing an over optimistic $\text{MMOM}$ result, although we believe that two MOM values with opposite signs can be considered better than with the same sign. 
To avoid this problem, which we call the cancelling of error problem, we also define the following Root mean square MOM (RMOM) criterion: 
\begin{equation} \label{Eq:RMOM}
\text{RMOM} = \sqrt{ \frac{1}{m}\sum_{i=1}^m \{ \text{MOM}(\bm{s}_{0i}) \} ^2 }. 
\end{equation}

We choose the prediction locations of a regular grid in the observation region, so the value of MLOE, MMOM, and RMOM can describe the average prediction performance
over the whole observation region. For instance, when the observation region is $[0, 1]^2$, the prediction locations can be $(i/5, j/5)$ for $i, j = 1, 2, 3, 4$. 
The MLOE describes the average efficiency loss of the prediction when the approximated covariance model is used instead of the true one, 
whereas the MMOM describes the average misspecification between the computed and true MSEs. 
The RMOM describes the degree of deviance of the misspecification between the computed and true MSEs from zero. 

In the case where the approximated model is an estimated model, \cite{stein1999interpolation} proposed alternative estimations for the MSE term $\text{E}_t \{ e_a^2(\bm{s}_0) \}$ in \eqref{Eq:LOE_one_point} and \eqref{Eq:MOM_one_point}. Let the random field $Z(\bm{s})$ follow a parameteric model, where the unknown parameter is $\bm{\theta}$. Denote the true value of this parameter by $\bm{\theta}_0$ and the estimated value by $\hat{\bm{\theta}} = \hat{\bm{\theta}}(\bm{Z})$. Here the approximated model is related to a random variable $\hat{\bm{\theta}}$, so the $\text{E}_t \{ e_a^2(\bm{s}_0) \}$ and $\text{E}_a \{ e_a^2(\bm{s}_0) \}$ terms in the LOE and MOM definitions are subsituted by their corresponding estimations, such as the plug-in estimation. Denote by $\text{E}_{\bm{\theta}}$, $\text{Var}_{\bm{\theta}}$, and $\text{Cov}_{\bm{\theta}}$ the expectation, variance, and covariance computed using the parametric model with parameter $\bm{\theta}$, respectively; $\hat{Z}_{\bm{\theta}}(\bm{s}_0)$ the kriging predictor under the parameter $\bm{\theta}$; $e_{\bm{\theta}}(\bm{s}_0) = \hat{Z}_{\bm{\theta}}(\bm{s}_0) - Z(\bm{s}_0)$ the prediction error; $K_{\bm{\theta}} = \text{Cov}_{\bm{\theta}} \{ \bm{Z}, \bm{Z}^\top \}$; $\bm{k}_{\bm{\theta}} = \text{Cov}_{\bm{\theta}} \{ \bm{Z}, Z(\bm{s}_0) \}$; $k_{0, \bm{\theta}} = \text{Var}_{\bm{\theta}} \{ Z(\bm{s}_0) \}$. Thus, 
$\text{E}_t \{ e_t^2(\bm{s}_0) \} = \text{E}_{\bm{\theta}_0} \{ e_{\bm{\theta}_0}^2(\bm{s}_0) \} =  k_{0, \bm{\theta}_0} - \bm{k}_{\bm{\theta}_0}^\top K_{\bm{\theta}_0}^{-1} \bm{k}_{\bm{\theta}_0}$. The $\text{E}_t \{ e_a^2(\bm{s}_0) \}$ and $\text{E}_a \{ e_a^2(\bm{s}_0) \}$ terms can be estimated by the following plug-in estimation: 
\begin{equation} \label{Eq:Etea2_Plugin}
\text{E}_t \{ e_a^2(\bm{s}_0) \} = \text{E}_{\bm{\theta}_0} \{ e_{\hat{\bm{\theta}}}^2(\bm{s}_0) \}
\approx \text{E}_{\bm{\theta}_0} \{ e_{\bm{\theta}}^2(\bm{s}_0) \} |_{\bm{\theta} = \hat{\bm{\theta}}} = 
k_{0, \bm{\theta}_0} - 2\bm{k}_{\bm{\theta}_0}^\top K_{\hat{\bm{\theta}}}^{-1} \bm{k}_{\hat{\bm{\theta}}} + \bm{k}_{\hat{\bm{\theta}}}^\top K_{\hat{\bm{\theta}}}^{-1} K_{\bm{\theta}_0} K_{\hat{\bm{\theta}}}^{-1} \bm{k}_{\hat{\bm{\theta}}}, 
\end{equation}
\begin{equation} \label{Eq:Eaea2_Plugin}
\text{E}_a \{ e_a^2(\bm{s}_0) \} 
\approx \text{E}_{\bm{\theta}} \{ e_{\bm{\theta}}^2(\bm{s}_0) \} |_{\bm{\theta} = \hat{\bm{\theta}}} = 
k_{0, \hat{\bm{\theta}}} - \bm{k}_{\hat{\bm{\theta}}}^\top K_{\hat{\bm{\theta}}}^{-1} \bm{k}_{\hat{\bm{\theta}}}. 
\end{equation}
\cite{stein1999interpolation} noted that, 
when $\{\bm{Z}^\top, Z(\bm{s}_0)\}^\top$ is Gaussian, the conditional distribution of $e_{\hat{\bm{\theta}}}(\bm{s}_0)$ given $\bm{Z} = \bm{z}$ is $N(e_{\hat{\bm{\theta}}(\bm{z})}(\bm{s}_0) - e_{\bm{\theta}_0}(\bm{s}_0), \text{E}_{\bm{\theta}_0} \{ e_{\bm{\theta}_0}^2(\bm{s}_0) \})$, so 
\begin{equation*}
\text{E}_{\bm{\theta}_0} \{ e_{\hat{\bm{\theta}}(\bm{Z})}^2(\bm{s}_0) | \bm{Z} = \bm{z} \} = \text{E}_{\bm{\theta}_0} \{ e_{\bm{\theta}_0}^2(\bm{s}_0) \} + \{ e_{\hat{\bm{\theta}}(\bm{z})}(\bm{s}_0) - e_{\bm{\theta}_0}(\bm{s}_0) \}^2. 
\end{equation*}
Therefore, the $\text{E}_t \{ e_a^2(\bm{s}_0) \}$ term can be estimated by 
\begin{equation} \label{Eq:Etea2_Stein}
\text{E}_t \{ e_a^2(\bm{s}_0) \} \approx \text{E}_{\bm{\theta}_0} \{ e_{\bm{\theta}_0}^2(\bm{s}_0) \} + \{ e_{\hat{\bm{\theta}}}(\bm{s}_0) - e_{\bm{\theta}_0}(\bm{s}_0) \}^2, 
\end{equation}
and $\text{E}_a \{ e_a^2(\bm{s}_0) \}$ is still estimated by \eqref{Eq:Eaea2_Plugin}. When our suggested criteria is computed using \eqref{Eq:Etea2_Plugin} and \eqref{Eq:Eaea2_Plugin}, we say that the criteria is computed using the plug-in method. When our criteria is computed using \eqref{Eq:Etea2_Stein} and \eqref{Eq:Eaea2_Plugin}, we say that the criteria is computed using Stein's method. 

In the plug-in method, the computed $\text{E}_{\bm{\theta}_0} \{ e_{\bm{\theta}}^2(\bm{s}_0) \} |_{\bm{\theta} = \hat{\bm{\theta}}}$ may be very slightly smaller than $\text{E}_{\bm{\theta}_0} \{ e_{\bm{\theta}_0}^2(\bm{s}_0) \} $, possibly due to round-off error. In the subsequent simulations of this article, the smallest value of the computed $\text{E}_{\bm{\theta}_0} \{ e_{\bm{\theta}}^2(\bm{s}_0) \} |_{\bm{\theta} = \hat{\bm{\theta}}} - \text{E}_{\bm{\theta}_0} \{ e_{\bm{\theta}_0}^2(\bm{s}_0) \} $ is $-6.3101\times 10^{-15}$. In this case, we estimate the $\text{E}_t \{ e_a^2(\bm{s}_0) \}$ term by $\text{E}_{\bm{\theta}_0} \{ e_{\bm{\theta}_0}^2(\bm{s}_0) \}$ instead, to keep LOE \eqref{Eq:LOE_one_point} nonnegative on all prediction locations. In Stein's method, the computed LOE is always nonnegative, which is better than the plug-in method. However, Stein's method did not consider the model misspecification case, so we particularly recommend Stein's method for the case when the parametric model is correctly specified, and recommend the plug-in method for more general cases, e.g., when the parametric model is misspecified. 
For simulations of this article, we will compute our suggested criteria using both of the methods. 

\cite{stein1999interpolation} also introduced a resampling method to better estimate $\text{E}_t \{ e_a^2(\bm{s}_0) \}$. This method first generates $n_r$ independent samples of $\bm{Z}$ and computes the estimate $\hat{\bm{\theta}}^{(1)}, \ldots, \hat{\bm{\theta}}^{(n_r)}$, then computes the kriging error terms $e_{\hat{\bm{\theta}}}(\bm{s}_0)$ and $e_{\bm{\theta}_0}(\bm{s}_0)$ for each sample, which are denoted by $e_{\hat{\bm{\theta}}^{(j)}}^{(j)}(\bm{s}_0)$ and $e_{\bm{\theta}_0}^{(j)}(\bm{s}_0)$, $j = 1, \ldots, n_r$, respectively. Since $\text{E}_{\bm{\theta}_0} \{ e_{\bm{\theta}_0}^2(\bm{s}_0) \}$ remains unchanged for resampling, we have the following estimation:
\begin{equation*}
\text{E}_t \{ e_a^2(\bm{s}_0) \} \approx \text{E}_{\bm{\theta}_0} \{ e_{\bm{\theta}_0}^2(\bm{s}_0) \} + \frac{1}{n_r} \sum_{j=1}^{n_r} \{ e_{\hat{\bm{\theta}}^{(j)}}^{(j)}(\bm{s}_0) - e_{\bm{\theta}_0}^{(j)}(\bm{s}_0) \}^2. 
\end{equation*}
In the simulation framework, such resampling method is equivalent to estimate $\text{E}_t \{ e_a^2(\bm{s}_0) \}$ using Stein's method \eqref{Eq:Etea2_Stein} with $n_r$ replicates and report the mean value as the final result. When the number of replicates of the simulation is large enough, the increment of samples with a price of more computational burdens may not be necessary. Therefore, we will not perform this resampling in our simulation. 


The computation of MLOE, MMOM, and RMOM criteria involves the inversion of the covariance matrix. In the simulation of Section \ref{sec:experimental_results}, where the number of observations is $n=3,600$, the computational times for these criteria are acceptable. If the direct computation of these criteria is not available due to the data size, one can adopt a matrix compression method, such as the TLR \citep{akbudak2017tile, abdulah2018tile}. This compression can provide an approximation of the covariance matrix and save computational time. 

\section{Simulation on the Validity of the Suggested Criteria} \label{Sec:Simulation_Misspecified}

We perform a numerical simulation to illustrate the validity and sensitivity of the suggested criteria, compared with the popular Mean Square Prediction Error (MSPE) criterion and the Kullback-Leibler Divergence criterion, which can be deduced by the logarithmic score introduced by \cite{gneiting2007strictly}. 
Similar to the settings in \cite{stein1999interpolation}, we focus on the case where the covariance model is misspecified. 

In this simulation, we consider a zero-mean stationary Gaussian random field $\{ Z(\bm{s}), \bm{s} \in [0, 1]^2 \}$ with Mat\'{e}rn covariance function \eqref{Eq:Matern_Cov}. 
We set the true covariance model as the exponential model, with covariance function $C(h; \bm{\theta}=(\sigma^2, \alpha, 0.5)^\top)$, and consider two cases of model misspecification. 
In the first case, the covariance model is correctly specified, but the parameters $\sigma^2$ and $\alpha$ are misspecified as their maximum likelihood estimate. 
Under this kind of misspecification, the corresponding kriging prediction is called `empirical best linear unbiased prediction' (EBLUP); the EBLUP does not significantly affect the prediction efficiency, according to the intuition and simulation results in the literature \citep{stein1999interpolation}. 
In the second case, the covariance model is misspecified as a smoother (Whittle) covariance model plus a nugget effect term. In this case, the misspecified covariance function is $C(h; \bm{\theta}=(\sigma^2, \alpha, 1.0)^\top)+\tau^2 I_{h=0}(h)$, where $\tau^2$ is the nugget variance and $I_{h=0}$ is the indicator function. 
In the \textsf{R} package $\texttt{fields}$, the function for fitting a covariance model $\texttt{MLESpatialProcess()}$ chooses the smoothness parameter $\nu=1$ as the default value, which is smoother than the common setting $\nu=0.5$. This motivates us to investigate the loss of prediction efficiency for this case. 

The observation locations are set to be $\bm{s}_{r, l} = n^{-1/2} (r-0.5+\tilde{U}_{r, l}, l-0.5+\tilde{V}_{r, l})$, where $n$ is the number of observations, $\tilde{U}_{r, l}$ and $\tilde{V}_{r, l}$ are $i.i.d.$ samples from the uniform distribution $U[-0.4, 0.4]$. 
By ordering $r, l$ lexicographically, these locations are also denoted by $\bm{s}_1, \ldots, \bm{s}_n$. 
We take $n = 12^2, 24^2$, or $48^2$. 
The true covariance function is set as $C(h; \bm{\theta}=(\sigma^2, \alpha, 0.5)^\top)$, where $\sigma^2=1$, $\alpha=0.2/(-\log (0.05))$, such that the true effective range of the model is $0.2$. 
For each parameter setting, we generate $100$ independent replications from the random field with the true covariance model at the same observation locations. First, we compute the MLE for the correctly specified and misspecified covariance models, then, we compute the plug-in kriging predictions for both covariance functions at the point $(i/5, j/5)$, for $i, j = 1, 2, 3, 4$, 
denoted by $\bm{s}_{01}, \ldots, \bm{s}_{0p}$ for $p=16$, 
and compare the results with the kriging results from the exact covariance functions. Lastly, the performance of the prediction is comparatively assessed, using our suggested MLOE \eqref{Eq:MLOE}, MMOM \eqref{Eq:MMOM}, and RMOM \eqref{Eq:RMOM} computed by the plug-in method and Stein's method, as well as MSPE in \eqref{Eq:Def_MSPE} and the Kullback-Leibler Divergence (K-L Divergence) criterion. As we have discussed in Section \ref{sec:method}, the plug-in method is more suitable for computing our suggested criteria in this simulation of model misspecification. 

The K-L divergence has been used to assess the estimation performance by comparing the approximated likelihood to the exact one \citep{huang2018hierarchical}. For predictions, we need to compare two predictive distributions. 
Let $Q_t$ be the distribution of $\bm{Z}_p := \left\{ Z(\bm{s}_{01}), \ldots, Z(\bm{s}_{0m})\right\}^\top$, conditional to the observations $\bm{Z} = \left\{ Z(\bm{s}_1), \ldots, Z(\bm{s}_n)\right\}^\top$, computed using the true model, and $Q_a$ be the computed distribution of $\bm{Z}_p$ conditional to $\bm{Z}$ using the approximated model. Denoting these two conditional distributions by $\{ \bm{Z}_p | \bm{Z} \}_t$ and $\{ \bm{Z}_p | \bm{Z} \}_a$, respectively, then the Kullback-Leibler Divergence is denoted by
\begin{equation*}
D_{KL}(Q_t \| Q_a) = D_{KL}\left( \{ \bm{Z}_p | \bm{Z} \}_t \| \{ \bm{Z}_p | \bm{Z} \}_a \right) 
= \int \log \left\{ \frac{q_t(\bm{Z}_p | \bm{Z})}{q_a(\bm{Z}_p | \bm{Z})} \right\} q_t(\bm{Z}_p | \bm{Z}) \text{d}\bm{Z}_p, 
\end{equation*}
where $q_t$ and $q_a$ are the conditional distributions corresponding to the true and the approximated model, respectively. 
When $Q_t \sim N(\bm{\mu}_{Q_t}, \bm{\Sigma}_{Q_t})$ and $Q_a \sim N(\bm{\mu}_{Q_a}, \bm{\Sigma}_{Q_a})$, the K-L divergence between these two multivariate Gaussian distribution satisfies 
\begin{equation} \label{Eq:K-L_Criterion}
\resizebox{\textwidth}{!}{
$D_{KL}(Q_t \| Q_a) = 
\frac{1}{2} \left\{ \text{trace}(\bm{\Sigma}_{Q_a}^{-1} \bm{\Sigma}_{Q_t}) - \log \det (\bm{\Sigma}_{Q_a}^{-1} \bm{\Sigma}_{Q_t}) + (\bm{\mu}_{Q_a} - \bm{\mu}_{Q_t})^\top \bm{\Sigma}_{Q_a}^{-1} (\bm{\mu}_{Q_a} - \bm{\mu}_{Q_t}) - m \right\}, $
}
\end{equation}
where $m$ is the dimension of $Q_t$ or $Q_a$. 

The K-L Divergence criterion comes from the logarithmic score criterion introduced by \cite{gneiting2007strictly}. Let $\bm{x} \in \mathbb{R}^m$ be the $m-$dimensional observed value and $\tilde{P}$ be the predicted distribution for this value, where $\tilde{P}$ is assumed to be only related to its mean $\bm{\mu}_{\tilde{P}}$ and covariance matrix $\bm{\Sigma}_{\tilde{P}}$. Then the scoring rule 
\begin{equation*}
S(\tilde{P}, \bm{x}) = -\log \det \bm{\Sigma}_{\tilde{P}} - (\bm{x} - \bm{\mu}_{\tilde{P}})^\top \bm{\Sigma}_{\tilde{P}}^{-1} (\bm{x} - \bm{\mu}_{\tilde{P}})
\end{equation*}
is strictly proper relative to the class of Gaussian measures and is equivalent to the logarithmic score \citep{gneiting2007strictly}. Therefore we call this scoring rule the logarithmic score. By \cite{gneiting2007strictly}, the divergence function for this rule is 
\begin{equation*} 
d(\tilde{P}, \tilde{Q}) = \text{trace}(\bm{\Sigma}_{\tilde{P}}^{-1} \bm{\Sigma}_{\tilde{Q}}) - \log \det (\bm{\Sigma}_{\tilde{P}}^{-1} \bm{\Sigma}_{\tilde{Q}}) + (\bm{\mu}_{\tilde{P}} - \bm{\mu}_{\tilde{Q}})^\top \bm{\Sigma}_{\tilde{P}}^{-1} (\bm{\mu}_{\tilde{P}} - \bm{\mu}_{\tilde{Q}}) - m, 
\end{equation*}
where $\tilde{P}$, $\tilde{Q}$ are $m$-dimensional distributions with mean $\bm{\mu}_{\tilde{P}}$, $\bm{\mu}_{\tilde{Q}}$ and covariance matrix $\bm{\Sigma}_{\tilde{P}}$, $\bm{\Sigma}_{\tilde{Q}}$, respectively. Here the divergence function of a scoring rule is defined by $d(\tilde{P}, \tilde{Q}) := S(\tilde{Q}, \tilde{Q}) - S(\tilde{P}, \tilde{Q})$, where $S(\tilde{P}, \tilde{Q}) = \int S(\tilde{P}, \bm{x}) \text{d}\tilde{Q}(\bm{x})$. The $d(\tilde{P}, \tilde{Q})$ can be considered as a logarithm score divergence criterion, which equals to two times of the K-L Divergence $D_{KL}(\tilde{Q} \| \tilde{P})$. 



\begin{figure}[htbp]
	\centering
	\includegraphics[width = \textwidth]{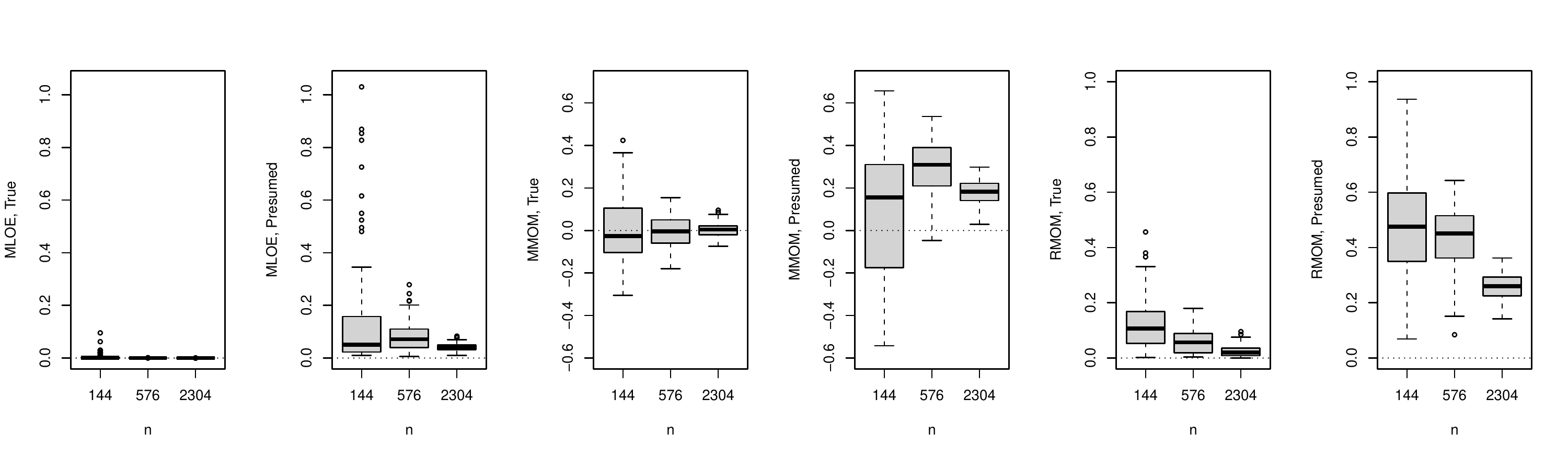}
	\caption{Boxplots of the MLOE, MMOM, and RMOM computed by the plug-in method, with respect to the number of observations $n$, when the covariance model is correctly specified as the exponential model (True) or misspecified as the Whittle model (Presumed). }
	\label{Fig:LOEMOM}
\end{figure}

\begin{figure}[htbp]
	\centering
	\includegraphics[width = \textwidth]{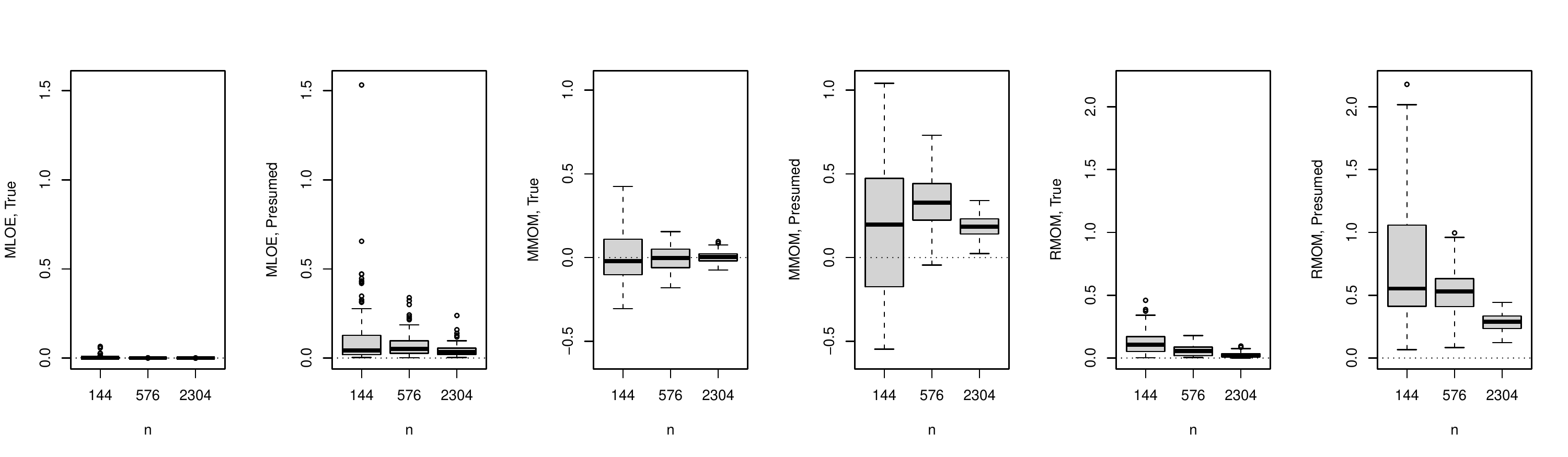}
	\caption{Boxplots of the MLOE, MMOM, and RMOM computed by Stein's method, with respect to the number of observations $n$, when the covariance model is correctly specified as the exponential model (True) or misspecified as the Whittle model (Presumed). }
	\label{Fig:LOEMOM_Stein}
\end{figure}

\begin{figure}[htbp]
	\centering
	\includegraphics[width = 0.8\textwidth]{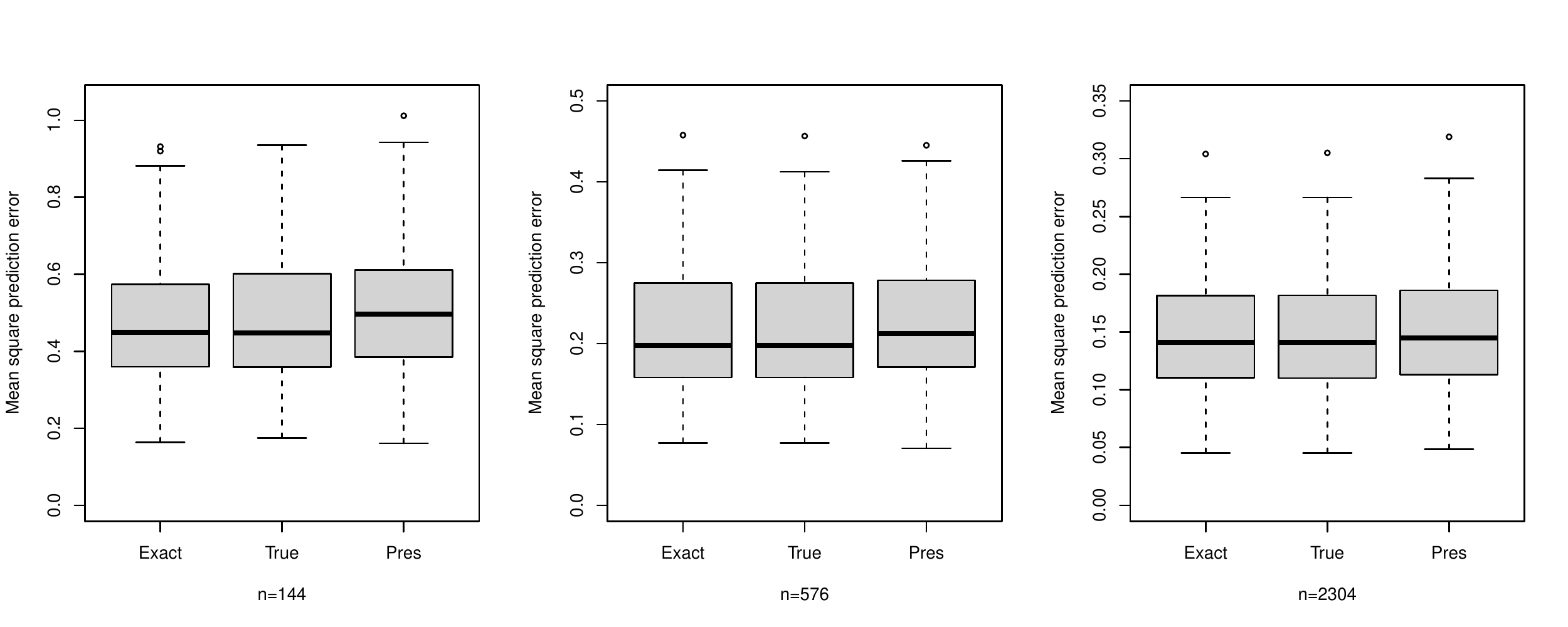}
	\caption{Boxplots of the MSPE for the predictions with respect to the exact model (Exact), plug-in prediction with correct covariance model (True), and the plug-in prediction with misspecified covariance model (Pres). }
	\label{Fig:MSPE}
\end{figure}

\begin{figure}[htbp]
	\centering
	\includegraphics[width = 0.6\textwidth]{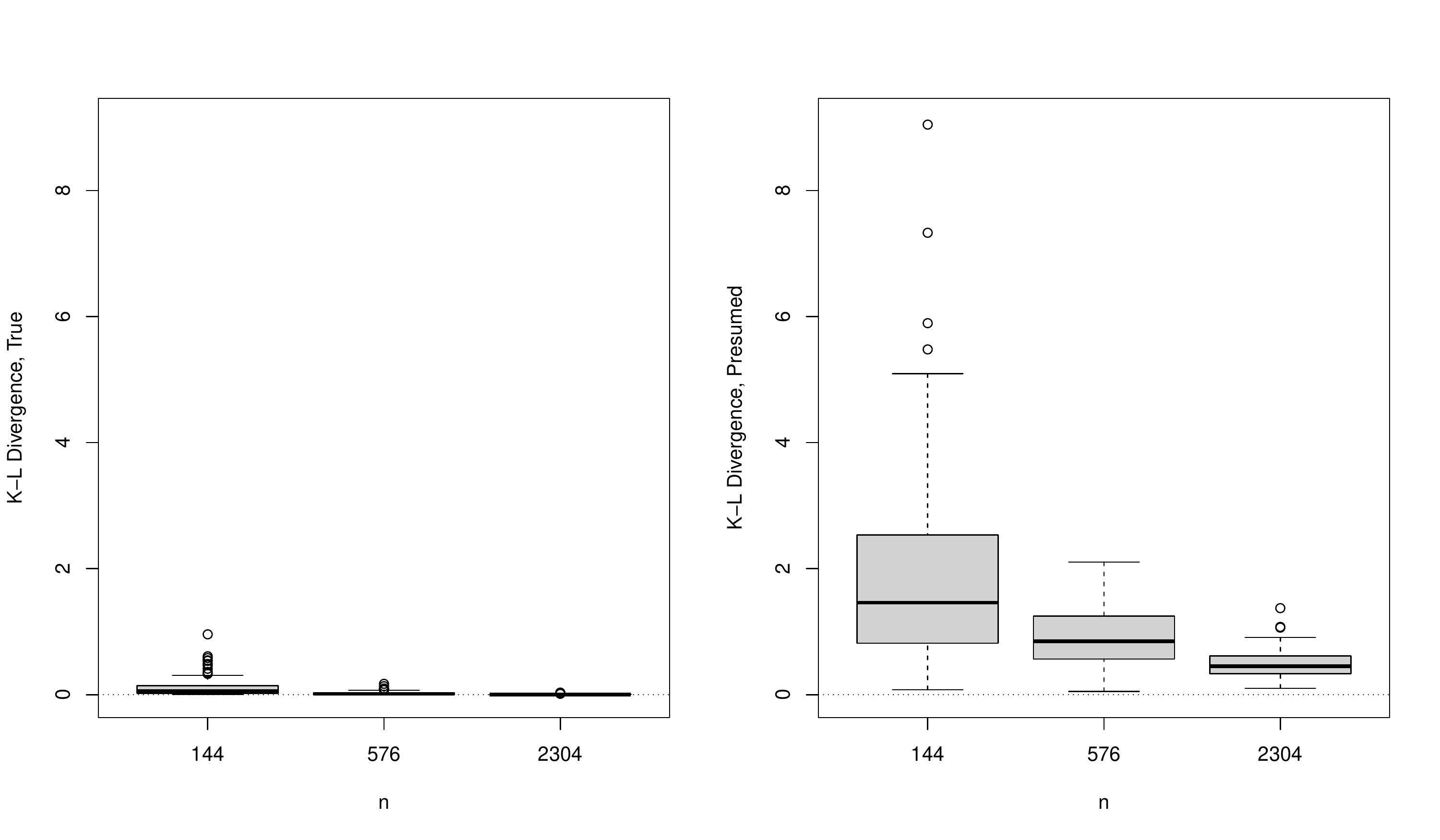}
	\caption{Boxplots of the K-L Divergence for the predictions with respect to the number of observations $n$, when the covariance model is correctly specified as the exponential model (True) or misspecified as the Whittle model (Presumed). }
	\label{Fig:K-L}
\end{figure}

The simulation results are shown in Figures \ref{Fig:LOEMOM}-\ref{Fig:K-L}. 
These figures show that, when the covariance model is correctly specified, the MLOE is very small in comparison with the exact model and has a decreasing trend when the number of observations $n$ increases. 
The MMOM is larger, but concentrates near zero and shrinks when $n$ increases. 
The RMOM also shrinks to zero when $n$ increases. 
This shows that the plug-in kriging prediction does not lead to a significant loss of prediction efficiency, which is in agreement with the intuition and the simulation results introduced in \cite{stein1999interpolation}. 
When the model is misspecified, the MLOE is clearly larger than that of the case where the model is correctly specified. 
The MMOM may likely have a mean value larger than zero, and the RMOM is larger than the correctly specified case. 
Thus, when a rougher covariance model is misspecified as a smoother model with a nugget effect, the plug-in prediction is suboptimal, and the MSE can be overestimated. 
In Figure \ref{Fig:MSPE}, the difference of boxplots between the case where the model is correctly specified or misspecified is not apparent, showing that our suggested MLOE, MMOM, and RMOM are more sensitive criteria for prediction accuracy. 
According to Figure \ref{Fig:K-L}, the K-L Divergence is also a sensitive criterion for prediction accuracy, measuring the information loss when the predicted distribution is approximated. However, it cannot provide more detailed information on the prediction for a spatial model that one may interested in, such as the efficiency loss of MSE when the model is approximated and the accuracy of the computed MSE. 
In conclusion, our suggested criteria are a valid, sensitive, and more informative tools to detect the loss of prediction efficiency caused by spatial model approximations in simulation studies.

\section{Simulation Experiments on the Tuning Parameters}
\label{sec:experimental_results}
As an application of the suggested MLOE, MMOM, and RMOM criteria in Section~\ref{sec:method}, 
we aim at assessing the performance of the TLR approximation method based
on these criteria. We define how to tune the TLR associated inputs based on the target data and the application requirements, 
using simulation experiments, which is the answer to the motivation of our study. 
All experiments being carried out in this section are conducted on a dual-socket 8-core Intel Sandy Bridge-based Xeon E5-2670 CPU 
running at 2.60GHz. 

\subsection{Simulation Settings}
\label{sec:datasets}
Here, we provide an outline of our simulation settings. 
Similar to the settings in \cite{sun2016statistically}, our simulation experiments
are performed on a set of synthetic datasets generated
using the built-in data generator tool in \exageostatr at irregular
locations in a 2D space (i.e., $\texttt{simulate\_data\_exact()}$ function). The generation process assumes a zero-mean stationary Gaussian
random field $\{Z(\bm{s}), \bm{s}\in [0, 1]^2 \}$. 
The observation locations $\bm{s}_1, \ldots, \bm{s}_n$ are generated by the same settings as those detailed in Section \ref{Sec:Simulation_Misspecified}. 
Given the set of $n$ locations, the covariance matrix $\Sigma$ is constructed using
the Mat\'{e}rn covariance function. 
  
  
  

The simulation is to illustrate the effectiveness of using the TLR approximation
method for the MLE estimation. The assessments include the total
execution time, estimation accuracy, and prediction accuracy. 
Instead of the MSPE criterion in \cite{abdulah2018tile}, 
here the prediction accuracy 
is investigated by MLOE, MMOM, and RMOM, using the plug-in method and Stein's method stated in Section \ref{sec:method}, 
which the effectiveness and the sensitivity have been shown in Section \ref{Sec:Simulation_Misspecified}. 
The assessment includes the kriging performance obtained
by using the estimated parameters to predict unknown sets of values at various specific locations.
Unless otherwise specified, our suggested criteria are computed by the plug-in method. Results of Stein's method are shown in Tables \ref{Table:nb_estimate_full_stein} - \ref{Table:predsupp_full_stein} in Supplementary Material, indicating that all conclusions drawn from the plug-in method and Stein's method are consistent. 
All the symbols in this section follow the abbreviations illustrated in Table~\ref{table:nonlin}.

In the simulation experiments conducted by \cite{abdulah2018tile}, both the estimation
and the prediction accuracy of the TLR method were shown by a set of
boxplots representing the estimation accuracy of different model parameters 
and the MSPE, and compared with the exact method. The
simulation performed in \cite{abdulah2018tile} also assessed the impact of using different TLR accuracy levels $tlr\_acc$ for
both the accuracy and the execution time. Here, we use our suggested criteria and not only consider $tlr\_acc$, but also
consider the impact of other tuning parameters, i.e., tile size, maximum rank, and optimization
tolerance to the overall execution time, estimation accuracy, and prediction accuracy. Moreover,
we consider two different smoothness levels of the underlying random field, i.e., $\nu=0.5$ and
$1$, whereas the simulations of \cite{abdulah2018tile} only considered $\nu=0.5$.

All the experiments in this section use spatial data where the number of locations is $n=3600$. 
For the true values of the parameters in \eqref{Eq:Matern_Cov}, we consider $\sigma^2=1$,
$\nu = 0.5$ or $1$, and $\alpha$ is chosen such that the effective range of the model can be
$h_{\text{eff}}=0.2, 0.4, 0.8$, or $1.6$. First, a set of the following experiments aims at comparing
the performance of the TLR approximation under different tile size $nb$ with suitable value of
maximum rank $tlr\_max\_rank$, while the other tuning parameters $tlr\_acc$ and
$opt\_tol$ are fixed at a moderate value which does not affect the estimation accuracy of
the approximations. Second, we compare the performance under different
accuracy levels $tlr\_acc$ and optimization tolerance $opt\_tol$, where the tile size $nb$ 
and maximum rank $tlr\_max\_rank$ are fixed at the suggested value obtained in the previous step. 
The reason for adopting these two steps is, according to \cite{abdulah2018tile}, that $nb$ and $tlr\_max\_rank$ mainly affect the computational time, whereas $tlr\_acc$ and $opt\_tol$ mainly affect the prediction efficiency. 
Recall that a larger $tlr\_acc$ corresponds to a coarser tile approximation, so the maximum rank necessary for the approximation is smaller. Thus the value of $tlr\_max\_rank$ does not directly affect the prediction efficiency; it could affect the efficiency via different $tlr\_acc$.

\subsection{Performance using Different Tile Sizes}
 \label{Sec:tlr+maxrank}
The parallel TLR approximation computation depends on dividing the matrix into a set of tiles where
the tile size is $nb \times nb$. Here, $nb$ should be tuned in different hardware platforms to obtain the best performance that corresponds to the trade-off between the arithmetic intensity and the degree of parallelism. 
We illustrate the performance and accuracy using different values of $nb$, i.e.,  $nb=400, 450, 600$, and $900$.  We fix $tlr\_acc=10^{-9}$ and $opt\_tol=10^{-6}$ since these values have little impact on the estimation performance. The $tlr\_max\_rank$ is fixed to the smallest feasible value for TLR computation obtained before the simulation. The $tlr\_max\_rank$ actually affects the memory allocation process and communication cost in case of distributed memory systems. A value of $tlr\_max\_rank$ that is too large can slow down the computation due to the unnecessary allocations, whereas a too small value may cause the failure of the SVD approximation of each off-diagonal tile. 
Thus, for each value of $nb$, we try to compute TLR approximations for $tlr\_max\_rank=10, 20, \ldots$, until the value of $tlr\_max\_rank$ can make the approximation feasible for all replicates.

For each parameter settings, we generate $100$ independent replicates of the observed random field. The
synthetic datasets are generated using the $\texttt{simulate\_data\_exact()}$ function in the \exageostatr package. The estimation performed uses both the exact and TLR methods, by the $\texttt{exact\_mle()}$ and the $\texttt{tlr\_mle()}$ functions in the same package, respectively, and estimate both the execution time and the estimation accuracy of each method, for different $nb$ values. The last step is 
to compute the MLOE, MMOM, and RMOM on prediction locations $(i/5, j/5)$, where $i, j = 1, 2, 3, 4$. The  
prediction performance is then evaluated  by the mean and standard deviations for both the values of our criteria. In our estimation, the value of $\nu$ is fixed at its true value and the optimization bound for estimating $\sigma^2$ and $\alpha$ is $[0.01, 5]$. 
The optimization tolerance of the exact MLE is set as $10^{-9}$ in order to get more accurate estimation results for comparison.

\begin{table}[htbp]
  \centering
  \caption{Smallest $tlr\_max\_rank$ that makes the TLR approximation applicable to different values of $nb$, and the parameters of the Mat\'{e}rn covariance. The number of locations is $n=3600$. } 
  \begin{tabular}{|c|c|cccc|}
  \hline 
  \multirow{2}{*}{$\nu$} & \multirow{2}{*}{Eff.range} & \multicolumn{4}{c|}{Tile size ($nb$)} \\
  \cline{3-6} 
      &    & 400 & 450 & 600 & 900 \\
  \hline 
  \multirow{4}{*}{$0.5$} & 0.2 & 260 & 210 & 310 & 270 \\
      & 0.4 & 250 & 210 & 310 & 270 \\
      & 0.8 & 250 & 210 & 300 & 260 \\
      & 1.6 & 250 & 210 & 300 & 260 \\ 
  \hline 
  \multirow{4}{*}{$1.0$} & 0.2 & 220 & 170 & 250 & 210 \\
      & 0.4 & 220 & 170 & 250 & 210 \\
      & 0.8 & 210 & 170 & 250 & 210 \\
      & 1.6 & 210 & 180 & 250 & 210 \\ 
  \hline 
  \end{tabular}
  \label{Table:lts_smallest_maxrank}
\end{table}

Selecting the smallest  $tlr\_max\_rank$ value for each tile size is important to obtain the best performance. Thus, 
we perform a set of experiments to select the $tlr\_max\_rank$ value
corresponding to each $nb$ when $n=3600$ (Table \ref{Table:lts_smallest_maxrank}).  The reported values show that the feasible $tlr\_max\_rank$ does not simply increase when the tile size $nb$ increases.  In fact, when the number of tiles is divisible by the number of underlying CPUs, i.e., $nb=450$ or $900$, the maximum rank of each tile $tlr\_max\_rank$ is relatively small compared to the $nb$. 
The required $tlr\_max\_rank$ is significantly smaller when the model is relatively smoother. Thus, when the number of locations is $n=3600$, we recommend to choose the $tlr\_max\_rank$ as the largest values shown in Table \ref{Table:lts_smallest_maxrank} for the corresponding values of $\nu$ and $nb$. 

For the MLE and different TLR approximations, we only show typical results with $\nu=0.5$, $h_{\text{eff}}=0.2$ and $\nu=1$, $h_{\text{eff}}=1.6$, shown in Table \ref{Table:nb}. 
This table shows that the TLR approximation has a similar estimation and prediction performance for different tile sizes $nb$, 
whereas the fastest computational time is obtained when $nb=450$. 

\begin{figure}[htbp]
	\centering
	\includegraphics[width=\textwidth]{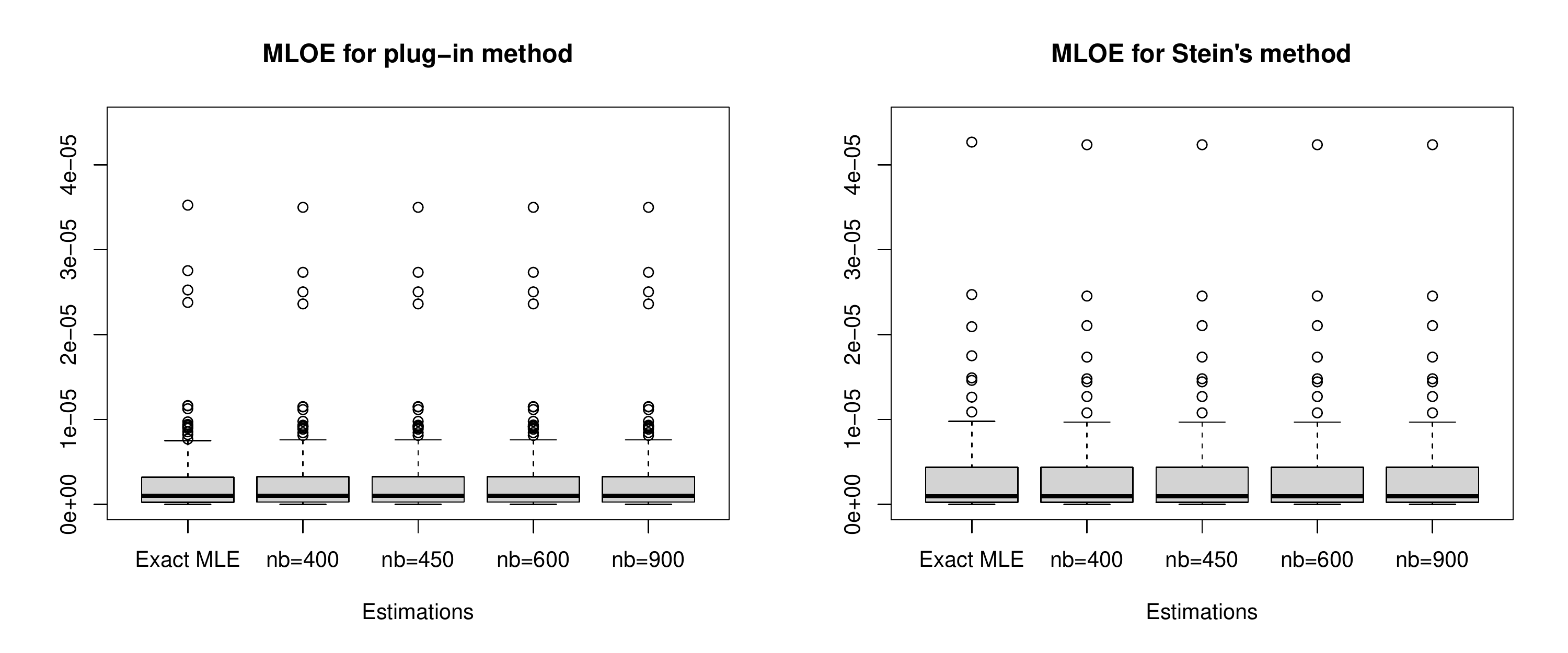} 
	\caption{Boxplots of the MLOE corresponding to MLE and TLR approximation estimates for different values of $nb$, where $\nu=0.5$ and $h_{\text{eff}} = 0.2$. } 
	\label{Fig:Example_MLOE} 
\end{figure}

\begin{figure}[htbp]
	\centering
	\includegraphics[width=\textwidth]{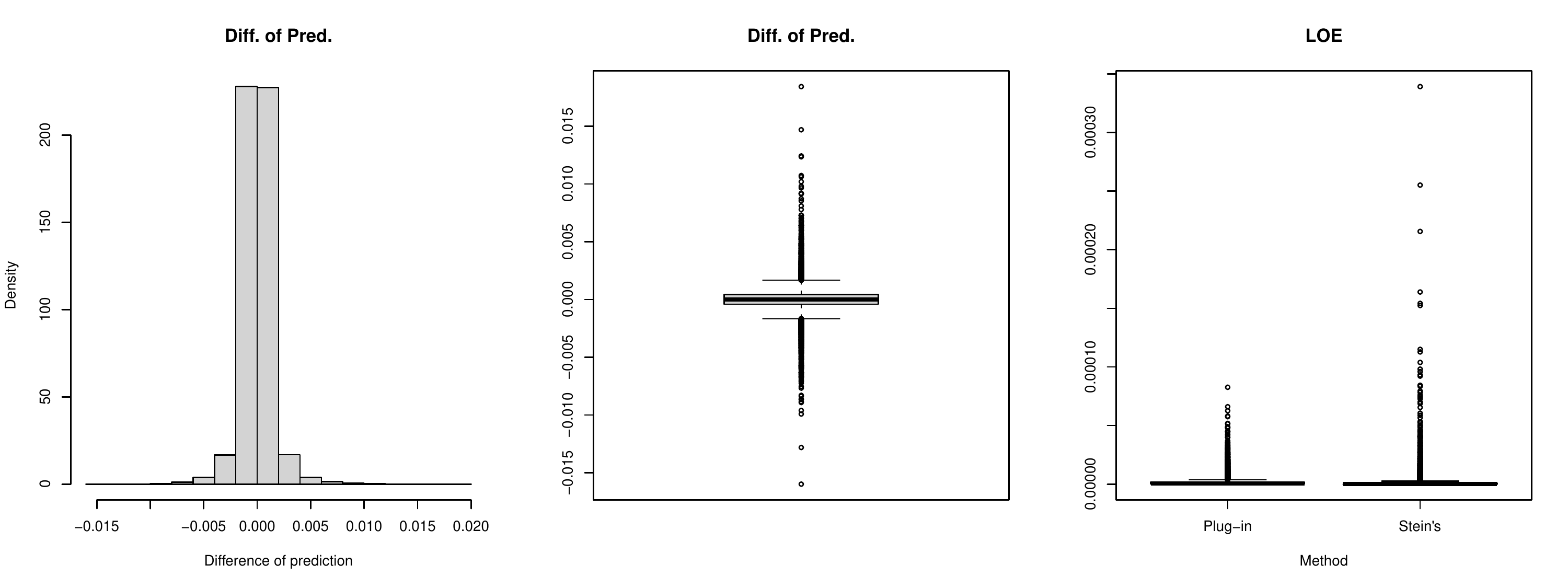} 
	\caption{Results of the difference of prediction results (Diff. of Pred.) and the LOE for an illustrating example of the prediction with misspecification. The left figure is the histogram of the difference of prediction; The middle is the boxplot of the difference of prediction; The right is the boxplot of the LOE computed by different methods. } 
	\label{Fig:Example} 
\end{figure}

\begin{Remark}
In Table \ref{Table:nb}, the standard deviation of the MLOE is larger than the corresponding mean. Figure \ref{Fig:Example_MLOE} shows the typical case of boxplots for the MLOE in the simulation, indicating that the distribution of MLOE is skewed to the right, causing a larger standard deviation. 

The main reason for the larger standard deviation is, when a normal distributed fluctuation is introduced in the model parameters, the difference of kriging prediction results between the original model and the fluctuated model may have a heavy-tailed distribution. We have run a simple illustrting example to show this. Consider a stationary Gaussian random field $Z(\bm{s})$ with exponential covariance function $C(h; \bm{\theta}) =  \sigma^2 \exp (-h/\alpha)$, where the observation locations are $\bm{s}_1 = (0, 0), \bm{s}_2 = (0, 1), \bm{s}_3 = (1, 0), \bm{s}_4 = (1, 1)$, and the prediction location is $\bm{s}_0 = (0.5, 0.5)$. We generate $10,000$ replicates of the observations, where the parameter $\bm{\theta} = (\sigma^2, \alpha)$ has true value $\sigma_0^2 = 1$, $\alpha_0^2 = 0.1$. In each replicate, the presumed values of $\sigma^2$ and $\alpha$ are independently drawn from normal distributions $N(1, 0.01^2)$ and $N(0.1, 0.01^2)$. We computed the difference of kriging prediction $\hat{Z}_{\bm{\theta}} (\bm{s}_0) - \hat{Z}_{\bm{\theta}_0} (\bm{s}_0)$ and $\text{LOE}(\bm{s}_0)$ for plug-in and Stein's methods, where $\bm{\theta}$ is the presumed value of parameters. Results are shown in Figure \ref{Fig:Example}, indicating that the difference $\hat{Z}_{\bm{\theta}} (\bm{s}_0) - \hat{Z}_{\bm{\theta}_0} (\bm{s}_0)$ follows a heavy-tailed distribution and $\text{LOE}(\bm{s}_0)$ is skewed to the right. The mean and standard deviation of $\text{LOE}(\bm{s}_0)$ are $1.8323 \times 10^{-6}$ and $4.2347 \times 10^{-6}$ (Plug-in method) or $1.8647 \times 10^{-6}$ and $8.1274 \times 10^{-6}$ (Stein's method), respectively. Note that for Stein's method, $\text{LOE}(\bm{s}_0) = \{ \hat{Z}_{\bm{\theta}} (\bm{s}_0) - \hat{Z}_{\bm{\theta}_0} (\bm{s}_0) \}^2 / \text{E}_{\bm{\theta}_0} \{ e_{\bm{\theta}_0}^2(\bm{s}_0) \}$, so the right-skewed distribution of the LOE comes from a square of a heavy-tailed distribution. The heavy-tailed distribution for the difference of kriging prediction results also appears in our simulation on different TLR tuning parameters. 

Although one can improve the accuracy of MLOE using a resampling method, we do not apply this in our simulation because the resampling method is equivalent to increasing the number of replications in the original simulation, as we have discussed in Section \ref{sec:method}. For the simple illustrating example stated above, one can run a resampling of $10,000$ cases and report the mean of the LOEs computed by Stein's method, e.g., $1.8647 \times 10^{-6}$, as the final LOE result. However, computing LOE without resampling for each replicate and reporting the mean and standard deviation, or the boxplot in Figure \ref{Fig:Example}, is more informative. 
\end{Remark}

\begin{table}[htbp]
	\centering
	\caption{Estimation and prediction performances of MLE and TLR approximation estimates for different values of $nb$. $\text{Bias}(\cdot)$ means the estimate of the parameter minus its true value, whereas the estimation time means the computational time of the corresponding estimation. The value of MLOE for all cases ($\nu=0.5$ and $\nu=1.0$) is multiplied by $10^6$. } 
	\resizebox{\textwidth}{!}{
	\begin{tabular}{|c|c|cccc|c|cccc|}
	\hline 
	\multirow{3}{*}{Mean (sd)} & \multicolumn{5}{c|}{$\nu=0.5$, $h_{\text{eff}}=0.2$} & \multicolumn{5}{c|}{$\nu=1.0$, $h_{\text{eff}}=1.6$} \\
	\cline{2-11}
		& \multirow{2}{*}{MLE} & \multicolumn{4}{c|}{TLR approximations ($nb$)} & \multirow{2}{*}{MLE} & \multicolumn{4}{c|}{TLR approximations ($nb$)} \\
	\cline{3-6} 
	\cline{8-11}
		&    & 400 & 450 & 600 & 900 &	    & 400 & 450 & 600 & 900 \\
	\hline
	\multirow{2}{*}{$\text{Bias} (\sigma^2)$} & $-$0.0080 & $-$0.0079 & $-$0.0079 & $-$0.0079 & $-$0.0079 & 0.0163 & 0.2173 & 0.2236 & 0.2210 & 0.2153 \\ 
	& (0.0908) & (0.0908) & (0.0908) & (0.0908) & (0.0908) & (0.6546) & (0.7339) & (0.7793) & (0.7624) & (0.7419) \\ 
    \cline{2-11}
	\multirow{2}{*}{$\text{Bias} (\alpha)$} & $-$0.0006 & $-$0.0006 & $-$0.0006 & $-$0.0006 & $-$0.0006 & $-$0.0144 & 0.0577 & 0.0577 & 0.0579 & 0.0571 \\ 
	& (0.0063) & (0.0063) & (0.0063) & (0.0063) & (0.0063) & (0.1186) & (0.1348) & (0.1394) & (0.1370) & (0.1357) \\ 
	\hline
	\multirow{2}{*}{MLOE $(\times 10^6)$} & 3.3945 & 3.3756 & 3.3756 & 3.3758 & 3.3756 & 0.0273 & 0.0109 & 0.0110 & 0.0109 & 0.0109 \\ 
	    & (5.9930) & (5.9474) & (5.9474) & (5.9477) & (5.9475) & (0.0669) & (0.0242) & (0.0243) & (0.0242) & (0.0241) \\ 
    \cline{2-11}
	\multirow{2}{*}{MMOM} & 0.0017 & 0.0011 & 0.0011 & 0.0011 & 0.0011 & 0.0014 & $-$0.1428 & $-$0.1428 & $-$0.1428 & $-$0.1428 \\ 
	& (0.0232) & (0.0232) & (0.0232) & (0.0232) & (0.0232) & (0.0227) & (0.0223) & (0.0222) & (0.0223) & (0.0223) \\ 
	\cline{2-11}
	\multirow{2}{*}{RMOM} & 0.0185 & 0.0185 & 0.0185 & 0.0185 & 0.0185 & 0.0182 & 0.1428 & 0.1428 & 0.1428 & 0.1428 \\ 
	& (0.0141) & (0.0140) & (0.0140) & (0.0140) & (0.0140) & (0.0135) & (0.0223) & (0.0222) & (0.0223) & (0.0223) \\ 
	\hline
	Estimation & 146.5 & 110.2 & 90.3 & 146.4 & 146.6 & 277.5 & 122.3 & 108.1 & 143.5 & 186.5 \\ 
	time (sec) & (20.2) & (15.3) & (13.5) & (21.7) & (19.5) & (75.6) & (35.6) & (31.6) & (46.6) & (63.0) \\ 
	\hline 
	\end{tabular}}
	\label{Table:nb}
\end{table}

\subsection{Performance using Different TLR Accuracy Levels } 
\label{Sec:assess-acc}

We investigate the effect of $tlr\_acc$ and $opt\_tol$ for the TLR approximations, where $nb=450$ and $tlr\_max\_rank$ is chosen from Table \ref{Table:lts_smallest_maxrank}. To compare the effect of different values of $tlr\_acc$, we fix $opt\_tol = 10^{-6}$ and choose $tlr\_acc = 10^{-5}$, $10^{-7}$, $10^{-9}$, or $10^{-11}$. We also compare the effect of different
$opt\_tol$ values; to do so, we fix $tlr\_acc=10^{-9}$ and choose $opt\_tol = 10^{-3}$, $10^{-6}$, $10^{-9}$, or $10^{-12}$.

The parameter settings and the simulation procedures are similar to those given in Section \ref{Sec:tlr+maxrank}. 
When $tlr\_acc = 10^{-11}$, the $tlr\_max\_rank$ value in Table \ref{Table:lts_smallest_maxrank} is not large enough. 
Thus, we use increased values of $tlr\_max\_rank$ in this case, namely, when $\nu=0.5$, we set $tlr\_max\_rank=270$ for $h_{\text{eff}}=0.2$ and $tlr\_max\_rank=260$ for other cases; when $\nu=1$, we set $tlr\_max\_rank=200$. 
We only provide the estimation and prediction performances for two typical cases, 
when $\nu=0.5$, $h_{\text{eff}}=0.2$, and when $\nu=1$, $h_{\text{eff}}=1.6$. 
Table \ref{Table:tlracc} shows the results obtained with different values of $tlr\_acc$, and Table \ref{Table:opttol} presents the results for different $opt\_tol$ values. 
For more detail, please refer to the Supplementary Material. 

\begin{table}[htbp]
    \centering
    \caption{Estimation and prediction performances of the exact MLE and TLR approximation estimates for different $tlr\_acc$ values. $\text{Bias}(\cdot)$ means the estimate of the parameter minus its true value, and the estimation time means the computational time of the corresponding estimation. The value of MLOE for all cases ($\nu=0.5$ and $\nu=1.0$) is multiplied by $10^6$. The missing part in the table (-) means that the result is not available, because the covariance matrix is numerically non positive-definite. } 
    \resizebox{\textwidth}{!}{
    \begin{tabular}{|c|c|cccc|c|cccc|}
	\hline 
	\multirow{3}{*}{Mean (sd)} & \multicolumn{5}{c|}{$\nu=0.5$, $h_{\text{eff}}=0.2$} & \multicolumn{5}{c|}{$\nu=1.0$, $h_{\text{eff}}=1.6$} \\
    \cline{2-11}
    	& \multirow{2}{*}{MLE} & \multicolumn{4}{c|}{TLR accuracy ($tlr\_acc$)} & \multirow{2}{*}{MLE} & \multicolumn{4}{c|}{TLR accuracy ($tlr\_acc$)} \\
    \cline{3-6} 
	\cline{8-11}
      &  & $10^{-5}$ & $10^{-7}$ & $10^{-9}$ & $10^{-11}$ &     & $10^{-5}$ & $10^{-7}$ & $10^{-9}$ & $10^{-11}$ \\
    \hline
    \multirow{2}{*}{$\text{Bias} (\sigma^2)$} & $-$0.0080 & $-$0.0023 & $-$0.0079 & $-$0.0079 & $-$0.0079 & 0.0163 & - & 0.3800 & 0.2236 & 0.2276 \\ 
    												& (0.0908) & (0.1095) & (0.0908) & (0.0908) & (0.0908) & (0.6546) & - & (0.3154) & (0.7793) & (0.7927) \\ 
    \cline{2-11}
    \multirow{2}{*}{$\text{Bias} (\alpha)$} & $-$0.0006 & $-$0.0002 & $-$0.0006 & $-$0.0006 & $-$0.0006 & $-$0.0144 & - & 0.1033 & 0.0577 & 0.0581 \\ 
    											& (0.0063) & (0.0077) & (0.0063) & (0.0063) & (0.0063) & (0.1186) & - & (0.0686) & (0.1394) & (0.1408) \\ 
    \hline
    \multirow{2}{*}{MLOE $(\times 10^6)$} & 3.3945 & 3.5691 & 3.3756 & 3.3756 & 3.3757 & 0.0273 & - & 0.0079 & 0.0110 & 0.0109 \\ 
                                                & (5.9931) & (6.4517) & (5.9476) & (5.9474) & (5.9474) & (0.0669) & - & (0.0167) & (0.0243) & (0.0242) \\ 
    \cline{2-11}
    \multirow{2}{*}{MMOM} & 0.0017 & 0.0009 & 0.0011 & 0.0011 & 0.0011 & 0.0014 & - & $-$0.1436 & $-$0.1428 & $-$0.1428 \\ 
                                                  & (0.0232) & (0.0233) & (0.0232) & (0.0232) & (0.0232) & (0.0227) & - & (0.0222) & (0.0222) & (0.0222) \\ 
    \cline{2-11}
    \multirow{2}{*}{RMOM} & 0.0185 & 0.0186 & 0.0185 & 0.0185 & 0.0185 & 0.0182 & - & 0.1436 & 0.1428 & 0.1428 \\ 
                                                  & (0.0141) & (0.0140) & (0.0140) & (0.0140) & (0.0140) & (0.0135) & - & (0.0222) & (0.0222) & (0.0222) \\ 
    \hline 
    Estimation  & 168.1 & 69.8 & 77.8 & 90.4 & 112.1 & 274.3 & - & 62.7 & 106.7 & 111.6 \\ 
    time (sec)   & (22.3) & (11.9) & (9.5) & (13.5) & (15.5) & (74.6) & - & (26.7) & (31.2) & (33.7) \\ 
	\hline 
    \end{tabular}}
    \label{Table:tlracc}
\end{table} 

\begin{table}[htbp]
    \centering
    \caption{Estimation and prediction performances of the exact MLE and TLR approximation estimates for different $opt\_tol$ values. $\text{Bias}(\cdot)$ means the estimate of the parameter minus its true value, and the estimation time means the computational time of the corresponding estimation. The value of MLOE for all cases ($\nu=0.5$ and $\nu=1.0$) is multiplied by $10^6$. } 
    \resizebox{\textwidth}{!}{
	\begin{tabular}{|c|c|cccc|c|cccc|}
	\hline 
	\multirow{3}{*}{Mean (sd)} & \multicolumn{5}{c|}{$\nu=0.5$, $h_{\text{eff}}=0.2$} & \multicolumn{5}{c|}{$\nu=1.0$, $h_{\text{eff}}=1.6$} \\
	\cline{2-11}
		& \multirow{2}{*}{MLE} & \multicolumn{4}{c|}{Optimization tolerance ($opt\_tol$)} & \multirow{2}{*}{MLE} & \multicolumn{4}{c|}{Optimization tolerance ($opt\_tol$)} \\
	\cline{3-6} 
	\cline{8-11}
	  &	& $10^{-3}$ & $10^{-6}$ & $10^{-9}$ & $10^{-12}$ &	& $10^{-3}$ & $10^{-6}$ & $10^{-9}$ & $10^{-12}$ \\
	\hline
	\multirow{2}{*}{$\text{Bias} (\sigma^2)$} & $-$0.0080 & 0.3654 & $-$0.0079 & $-$0.0079 & $-$0.0079 & 0.0163 & 0.3263 & 0.2236 & 0.2234 & 0.2234 \\  
	  & (0.0908) & (0.3017) & (0.0908) & (0.0908) & (0.0908) & (0.6546) & (0.4421) & (0.7793) & (0.7787) & (0.7787) \\ 
	\cline{2-11} 
	\multirow{2}{*}{$\text{Bias} (\alpha)$} & $-$0.0006 & 0.0253 & $-$0.0006 & $-$0.0006 & $-$0.0006 & $-$0.0144 & 0.0896 & 0.0577 & 0.0577 & 0.0577 \\ 
	  & (0.0063) & (0.0210) & (0.0063) & (0.0063) & (0.0063) & (0.1186) & (0.0904) & (0.1394) & (0.1393) & (0.1393) \\ 
	\hline
	\multirow{2}{*}{MLOE $(\times 10^6)$} & 3.3945 & 19.3107 & 3.3756 & 3.3756 & 3.3756 & 0.0273 & 0.0150 & 0.0110 & 0.0110 & 0.0110 \\ 
	  & (5.9931) & (13.9768) & (5.9474) & (5.9475) & (5.9475) & (0.0669) & (0.0708) & (0.0243) & (0.0243) & (0.0243) \\ 
	\cline{2-11}
	\multirow{2}{*}{MMOM} & 0.0017 & $-$0.0061 & 0.0011 & 0.0011 & 0.0011 & 0.0014 & $-$0.1432 & $-$0.1428 & $-$0.1428 & $-$0.1428 \\ 
	  & (0.0232) & (0.0234) & (0.0232) & (0.0232) & (0.0232) & (0.0227) & (0.0220) & (0.0222) & (0.0222) & (0.0222) \\ 
	\cline{2-11}
	\multirow{2}{*}{RMOM} & 0.0185 & 0.0196 & 0.0185 & 0.0185 & 0.0185 & 0.0182 & 0.1432 & 0.1428 & 0.1428 & 0.1428 \\ 
	  & (0.0141) & (0.0140) & (0.0140) & (0.0140) & (0.0140) & (0.0135) & (0.022) & (0.0222) & (0.0222) & (0.0222) \\ 
	\hline
	Estimation & 168.1 & 33.8 & 90.4 & 102.2 & 113.0 & 274.3 & 29.4 & 106.7 & 124.1 & 136.2 \\ 
	time (sec) & (22.3) & (13.6) & (13.5) & (13.2) & (13.0) & (74.6) & (21.7) & (31.2) & (34.8) & (36.0) \\ 
	\hline 
	\end{tabular}}
    \label{Table:opttol}
\end{table}

Tables \ref{Table:tlracc} and \ref{Table:opttol} indicate that the exact MLE and the TLR approximations can provide accurate prediction results since the small MLOE values suggest that the loss of prediction efficiency is very small. 
The MMOM results indicate that the computed MSEs are also accurate, except when $\nu=1$ and $h_{\text{eff}}=1.6$, which shows that the plug-in kriging based on TLR approximations may underestimate the prediction MSEs for a smoother random field with a larger effective range. 
The RMOM results exclude the canceling of error problem in the MMOM results. 
The plug-in kriging based on the exact MLE works well for all cases. 

Table \ref{Table:tlracc} shows that the TLR approximations give similar and relatively satisfactory performances of the estimation when $tlr\_acc \le 10^{-9}$, and that the prediction performs well when $tlr\_acc \le 10^{-7}$. 
The computational time increases when $tlr\_acc$ decreases, so we suggest $tlr\_acc = 10^{-9}$ for maintaining estimation performance and $tlr\_acc = 10^{-7}$ for maintaining prediction performance. 

Table \ref{Table:opttol} shows that the estimation performs relatively well when $opt\_tol \le 10^{-6}$. 
For prediction performances, the case of $opt\_tol=10^{-3}$ performs well enough, though the MLOE values are larger compared with other cases. 
So we suggest $opt\_tol = 10^{-6}$ for keeping estimation performances and $opt\_tol = 10^{-3}$ for keeping prediction performances because of the significantly faster computational speed in this case. 

To further investigate the impact of different combinations of $tlr\_acc$ and $opt\_tol$ for prediction performances, we also try the cases where $tlr\_acc$ can be $10^{-7}, 10^{-9}$ and $opt\_tol$ can be $10^{-3}, 10^{-6}$. 
Table \ref{Table:predsupp} shows that choosing $tlr\_acc=10^{-7}$ and $opt\_tol=10^{-3}$ can provide a faster computation without losing too much prediction efficiency; we therefore suggest to select $tlr\_acc=10^{-7}$ and $opt\_tol=10^{-3}$ for keeping the prediction performances. 

\begin{table}[htbp]
  \centering
  \caption{Prediction performance and the computational time for TLR approximations with different combinations of $tlr\_acc$ and $opt\_tol$. The estimation time means the computational time of the corresponding estimation. The value of MLOE for all cases ($\nu=0.5$ and $\nu=1.0$) is multiplied by $10^6$. } 
  \resizebox{\textwidth}{!}{
  \begin{tabular}{|c|cccc|cccc|}
  \hline 
  \multirow{2}{*}{Mean (sd)} & \multicolumn{4}{c|}{$(tlr\_acc, opt\_tol)$, $\nu=0.5$, $h_{\text{eff}}=0.2$} & \multicolumn{4}{c|}{$(tlr\_acc, opt\_tol)$, $\nu=1.0$, $h_{\text{eff}}=1.6$}\\
  \cline{2-9}
      & $(10^{-7}, 10^{-3})$ & $(10^{-9}, 10^{-3})$ & $(10^{-7}, 10^{-6})$ & $(10^{-9}, 10^{-6})$ & $(10^{-7}, 10^{-3})$ & $(10^{-9}, 10^{-3})$ & $(10^{-7}, 10^{-6})$ & $(10^{-9}, 10^{-6})$ \\
  \hline 
  \multirow{2}{*}{MLOE $(\times 10^6)$} & 19.0060 & 19.3107 & 3.3756 & 3.3756 & 0.0138 & 0.0150 & 0.0079 & 0.0110 \\ 
      & (13.6484) & (13.9768) & (5.9476) & (5.9474) & (0.0688) & (0.0708) & (0.0167) & (0.0243) \\ 
  \cline{2-9} 
  \multirow{2}{*}{MMOM} & $-$0.0062 & $-$0.0061 & 0.0011 & 0.0011 & $-$0.1436 & $-$0.1432 & $-$0.1436 & $-$0.1428 \\ 
      & (0.0232) & (0.0234) & (0.0232) & (0.0232) & (0.0219) & (0.0220) & (0.0222) & (0.0222) \\ 
      \cline{2-9} 
  \multirow{2}{*}{RMOM} & 0.0193 & 0.0196 & 0.0185 & 0.0185 & 0.1436 & 0.1432 & 0.1436 & 0.1428 \\ 
  & (0.0142) & (0.0140) & (0.0140) & (0.0140) & (0.0219) & (0.0220) & (0.0222) & (0.0222) \\ 
  \hline
  Estimation & 28.8 & 33.1 & 76.6 & 88.5 & 19.4 & 29.8 & 63.5 & 108.1 \\ 
  time (sec) & (11.4) & (13.3) & (9.4) & (13.2) & (8.8) & (22.1) & (27.1) & (31.3) \\ 
  \hline 
  \end{tabular}}
  \label{Table:predsupp} 
\end{table}

In conclusion, the TLR approximation method can significantly reduce the computational time and maintain the prediction efficiency. The only problematic aspect of the TLR method is that, when $\nu=1$ and the effective range is large, the prediction MSE may be underestimated. 
For tuning the inputs in the TLR approximation, we recommend a moderate value of $nb$ that makes the number of tiles divisible by the total number of CPUs, and a smallest feasible $tlr\_max\_rank$, which can be obtained by our simulations or by some simple trials. 
For instance, in Section \ref{sec:application}, we choose $nb$ according to the simulation in this section, such that the number of tiles remains the same for different $n$; $tlr\_max\_rank$ is determined by trials similar to the simulation. 
We suggest $tlr\_acc = 10^{-9}$, $opt\_tol = 10^{-6}$ for maintaining estimation performances; and suggest $tlr\_acc = 10^{-7}$, $opt\_tol = 10^{-3}$, when only the prediction performances are necessary to maintain. 

Our suggested MLOE, MMOM, and RMOM criteria can successfully assess the loss of spatial prediction efficiency of the TLR method with different tuning parameters. 
We have also succesfully detected the changes of the prediction efficiency for different $tlr\_acc$ and $opt\_tol$. 
For different $nb$, the criteria's values are similar, indicating that the $nb$ mainly affects the computational time, rather than the efficiency. 

\begin{Remark} \label{Remark:smoother}
Table \ref{Table:tlracc} shows that 
the TLR method with $tlr\_acc = 10^{-5}$ can maintain the prediction performance for the exponential covariance model, 
but cannot for the Whittle covariance model, 
suggesting that one may need a lower $tlr\_acc$ value for a smoother process. 
Thus, if the process is smoother than the Whittle covariance model and $tlr\_acc = 10^{-7}$ is not applicable, 
then one can choose smaller $tlr\_acc$ such as $10^{-9}$. 
\end{Remark}

\section{Application to Soil Moisture Data} \label{sec:application}

To show the effectiveness of our suggested TLR tuning parameter settings for real datasets, 
we compare the estimation and prediction performance of the TLR  approximation to the exact MLE for the soil moisture dataset, 
with a 64-bit 20-core Intel Xeon Gold 6248 CPU running at 2.50 GHz,
allowing the computation of the exact MLE by the \exageostatr framework. 
We use 4 nodes, each node has 16 underlying CPUs, 
so the number of tiles is divisible by the total number of CPUs. 

\begin{figure}[htbp]
    \centering
    \includegraphics[width=0.7\textwidth]{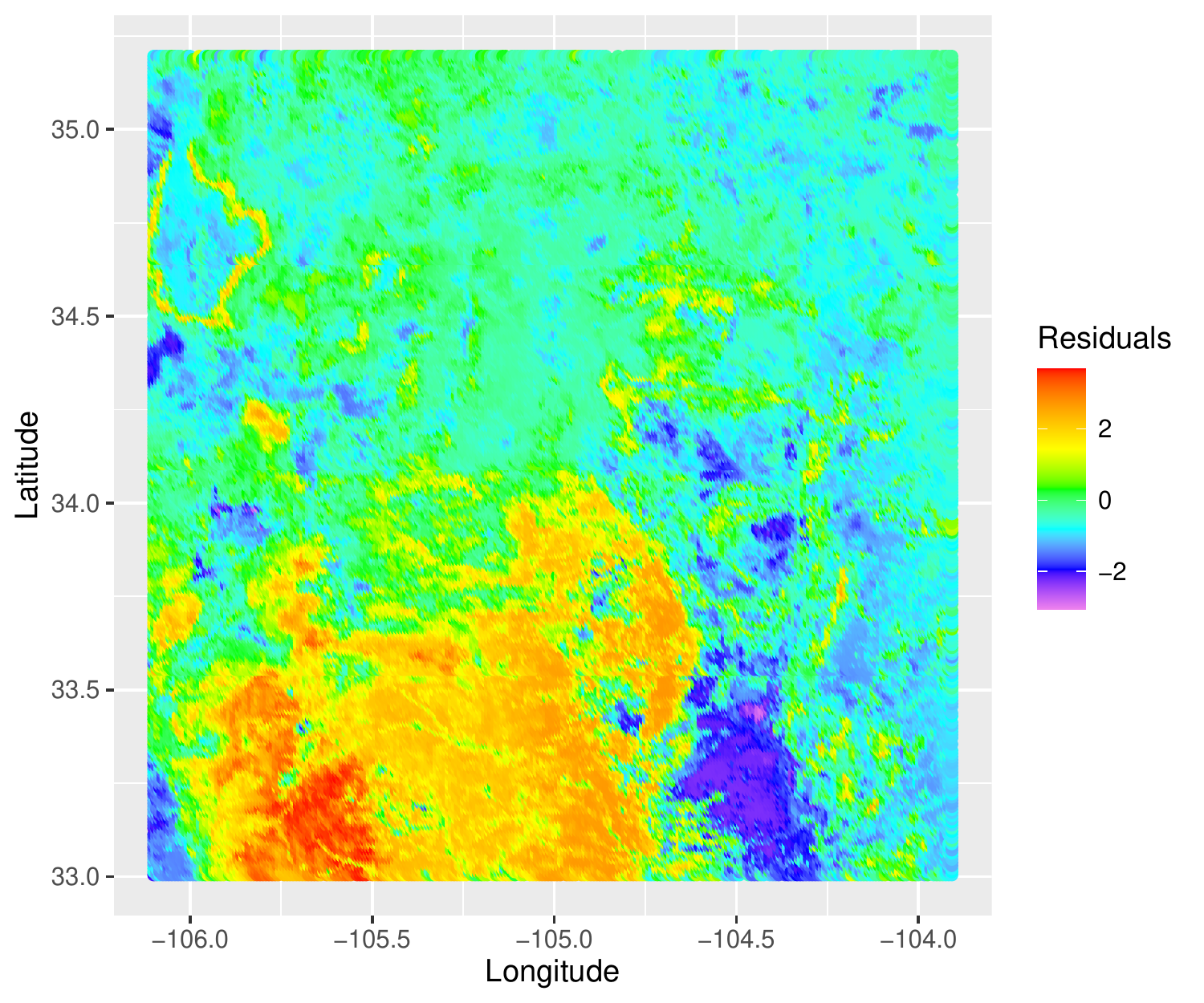} 
    \caption{Image plot of a subset of the soil moisture dataset residuals with $N=64,648$ for the real case study. } 
    \label{Fig:SoilData} 
\end{figure}

This dataset describes the daily soil moisture percentage at the top layer of the Mississippi basin, U.S., on January 1st, 2004, 
including the observation locations and the residuals of the fitted linear model in \cite{huang2018hierarchical}, and 
can be obtained from the website \url{https://ecrc.github.io/exageostat/md_docs__examples.html}, containing the example data of the \exageostat package. 
The full dataset consists of about 2 million locations, however,
we select a region of $N=64,648$ locations that can be considered as representative regions for the whole area. 
For our computational experiment, we consider a subset of this dataset, where the latitude and longitude of the locations lie within $[33.0, 35.2] \times [-106.1, -103.9]$, as shown in Figure \ref{Fig:SoilData}. 
We use the latitude and longitude as the coordinates of the observation locations in our computation. 

In this numerical experiment, we randomly choose $n=3600$, 14,400, 32,400, or 57,600 points for the estimation, and use the remaining points for assessing the prediction performance. 
For estimation, the smoothness parameter $\nu$ is either treated as unknown or fixed at $\nu=0.5$. 
The searching intervals for optimizing the likelihood function are $\sigma^2 \in [0.01, 5]$, $\alpha \in [0.01, 5]$, and $\nu \in [0.01, 5]$ for the unknown case. 
In TLR approximations, we choose $tlr\_acc = 10^{-9}$, $opt\_tol = 10^{-6}$ and $tlr\_acc = 10^{-7}$, $opt\_tol = 10^{-3}$, which are our recommendations for keeping the estimation and prediction performances, respectively. 
The $tlr\_max\_rank$ value is determined using the procedure presented in Section \ref{Sec:tlr+maxrank}. 
The settings of $nb$ and $tlr\_max\_rank$ are shown in Table \ref{Table:soil_lts_smallest_maxrank}, 
and results are given in Table \ref{Table:soil_result}. 
We also try $tlr\_acc=10^{-5}$, $opt\_tol = 10^{-3}$ or $10^{-6}$ in this experiment, but these parameter settings cannot work for $n=14,400$, $32,400$ and $57,600$, because the covariance matrix is numerically non positive-definite, similar as in Table \ref{Table:tlracc}. Therefore, we ignore the results with $tlr\_acc=10^{-5}$. 

\begin{table}[htb]
  \centering
  \caption{The value of $nb$ and the corresponding $tlr\_max\_rank$ used in the estimation of the soil moisture data. } 
  \begin{tabular}{|c|ccc|}
  \hline 
  \multirow{2}{*}{$n$} & \multirow{2}{*}{$nb$} & $tlr\_max\_rank$ & $tlr\_max\_rank$ \\
      &    & ($\nu$ unknown) & ($\nu$ known) \\
  \hline 
  3600 & 450 & 210 & 210 \\
  14,400 & 900 & 310 & 320 \\
  32,400 & 1350 & 490 & 500 \\
  57,600 & 1800 & 430 & 430 \\
  \hline 
  \end{tabular}
  \label{Table:soil_lts_smallest_maxrank}
\end{table}

\begin{table}[htb]
  \centering
  \caption{Estimation results, computational time and MSPE of the MLE and TLR estimation for soil moisture data, where $\nu$ is unknown or fixed at $0.5$. } 
  \resizebox{\textwidth}{!}{
  \begin{tabular}{|c|c|c|ccccc|cccc|}
  \hline 
  \multirow{2}{*}{$n$} & \multirow{2}{*}{$tlr\_acc$} & \multirow{2}{*}{$opt\_tol$} & \multicolumn{5}{c|}{$\nu$ is unknown} & \multicolumn{4}{c|}{$\nu$ is fixed at $0.5$} \\
  \cline{4-12}
  	&	&	& $\sigma^2$ & $\alpha$ & $\nu$ & Times (sec) & MSPE & $\sigma^2$ & $\alpha$ & Times (sec) & MSPE \\
  \hline 
  \multirow{3}{*}{3600} &  \multicolumn{2}{c|}{Exact MLE} & 1.2488 & 0.4590 & 0.2970 & 2038.1 & 0.2283 & 1.0970 & 0.1105 & 663.1 & 0.2335 \\ 
  \cline{2-12} 
      & $10^{-7}$ & $10^{-3}$ & 1.1163 & 0.3800 & 0.2971 & 83.5 & 0.2283 & 1.1842 & 0.1191 & 41.7 & 0.2337 \\ 
      & $10^{-9}$ & $10^{-6}$ & 1.2486 & 0.4587 & 0.2971 & 399.7 & 0.2283 & 1.0969 & 0.1106 & 142.9 & 0.2335 \\ 
  \hline 
  \multirow{3}{*}{14,400} &  \multicolumn{2}{c|}{Exact MLE} & 1.1412 & 0.2358 & 0.3566 & 13550.4 & 0.1461 & 1.0046 & 0.0864 & 8075.8 & 0.1457 \\ 
  \cline{2-12}
      & $10^{-7}$ & $10^{-3}$ & 0.8784 & 0.1740 & 0.3488 & 1331.5 & 0.1462 & 1.2210 & 0.1054 & 454.2 & 0.1458 \\ 
      & $10^{-9}$ & $10^{-6}$ & 1.1410 & 0.2356 & 0.3568 & 3248.5 & 0.1461 & 1.0046 & 0.0865 & 1863.5 & 0.1457 \\ 
  \hline 
  \multirow{3}{*}{32,400} & \multicolumn{2}{c|}{Exact MLE} & 1.0478 & 0.1263 & 0.4282 & 83197.3 & 0.1067 & 0.9800 & 0.0797 & 36503.7 & 0.1060 \\  
  \cline{2-12}
      & $10^{-7}$ & $10^{-3}$ & 1.4342 & 0.1898 & 0.4229 & 8687.8 & 0.1068 & 1.1183 & 0.0913 & 3183.4 & 0.1060 \\ 
      & $10^{-9}$ & $10^{-6}$ & 1.0475 & 0.1261 & 0.4285 & 25414.0 & 0.1067 & 0.9801 & 0.0798 & 14840.5 & 0.1060 \\ 
  \hline 
  \multirow{3}{*}{57,600} & \multicolumn{2}{c|}{Exact MLE} & 0.9870 & 0.0774 & 0.5066 & 286729.5 & 0.0811 & 0.9935 & 0.0805 & 184842.0 & 0.0812 \\ 
  \cline{2-12}
      & $10^{-7}$ & $10^{-3}$ & 1.1911 & 0.1002 & 0.4928 & 15807.4 & 0.0813 & 1.1260 & 0.0916 & 5680.1 & 0.0812 \\ 
      & $10^{-9}$ & $10^{-6}$ & 0.9868 & 0.0773 & 0.5071 & 74225.3 & 0.0811 & 0.9937 & 0.0806 & 21907.0 & 0.0812 \\ 
  \hline 
  \end{tabular}}
  \label{Table:soil_result}
\end{table}

Table \ref{Table:soil_result} indicates that, when $tlr\_acc = 10^{-9}$ and $opt\_tol = 10^{-6}$, the TLR approximation can provide parameter estimates that are very close to the exact MLE, with a significantly shorter computational time. 
The computational times are further shortened when $tlr\_acc = 10^{-7}$ and $opt\_tol = 10^{-3}$. 
In this case, the prediction performances are similar to the exact MLE, though the estimates are no longer similar. 
Thus, our proposed tuning parameter suggestions work well for the soil moisture dataset, showing that our suggested MLOE, MMOM, and RMOM criteria are successfully used to choose the tuning parameters for the TLR approximation. 


Besides the TLR method, we also compare the estimation and prediction performance for the soil moisture dataset with the composite likelihood method proposed by \cite{vecchia1988estimation} and implemented by \cite{guinness2020gpgp}, 
and the Gaussian predictive process method proposed by \cite{banerjee2008gaussian}. We still consider $n=3600$, 14,400, 32,400, 57,600 and use the same soil moisture dataset stated above. 

The composite likelihood approximates the log-likelihood function $\ell (\bm{\theta})$ by 
\begin{equation*}
\tilde{\ell}_m (\bm{\theta}) := \sum_{i=1}^n \log p(Z(\bm{s}_i)|Z(\bm{s}_{i1}), \ldots, Z(\bm{s}_{im})), 
\end{equation*}
where $\bm{s}_{i1}, \ldots, \bm{s}_{im}$ are $\min (i-1, m)$ locations that are nearest to $Z(\bm{s}_i)$, $p$ is the density of $Z(\bm{s}_i)$ conditional on the observations on these nearest locations. In this numerical experiment, we adopt the function  \texttt{vecchia\_meanzero\_loglik()} in the \texttt{GpGp} package \citep{guinness2020gpgp} to compute $\tilde{\ell}_m$, which employs a Fisher scoring algorithm introduced by \cite{guinness2019gaussian}. 
We use the function \texttt{constrOptim()} to compute the optimum. 
We set $m=20$ and choose the initial value for the optimization by $(\sigma^2, \alpha, \nu) = (1, 0.2, 0.5)$. 
Denote by $\bm{\hat{\theta}}_n = (\hat{\sigma}_n^2, \hat{\alpha}, \hat{\nu})$ the estimation results for different $n$, then we have 
$\bm{\hat{\theta}}_{3600} = (1.1716, 0.3778, 0.3057)$; $\bm{\hat{\theta}}_{14,400} = (1.0757, 0.2032, 0.3646)$; 
$\bm{\hat{\theta}}_{32,400} = (1.0243, 0.1171, 0.4359)$; 
$\bm{\hat{\theta}}_{57,600} = (0.9598, 0.0726, 0.5140)$. 
The MSPE results for $n=3600$, $14,400$, $32,400$, and $57,600$ are $0.2283, 0.1459, 0.1066$, and $0.0810$, respectively, and the computational times (in seconds) are $123.1, 410.6, 1112.7$, and $2805.1$, respectively. One can check from Table \ref{Table:soil_result} that, the TLR estimates for $tlr\_acc=10^{-9}$, $opt\_tol=10^{-6}$ are closer to the exact MLE results, compared to the composite likelihood estimates. Despite that the MSPE results of the TLR are slightly less competitive, it is clear that the TLR with our suggested tuning parameters for keeping the estimation performance reaches our goal for approximating the exact MLE estimation results, which serves our purpose better than the composite likelihood. Thus, with our suggested parameters, one can get more accurate information about the properties for the random field corresponding to the soil moisture dataset from a more accurate approximation of the MLE. Results for the case when $\nu$ is fixed at $0.5$ are similar and not reported here. 

Next, we compare the TLR with a typical low-rank based approximation method, the 
Gaussian predictive process (GPP) method proposed by \cite{banerjee2008gaussian}. 
For some predetermined knots $\bm{s}_1^\star, \ldots, \bm{s}_m^\star$, the GPP method approximates the observed value $Z(\bm{s})$ by its kriging prediction value with respect to the observations on the knots plus a nugget term. 
Denote the observations on the knots by $\bm{Z}^\star := \{ Z(\bm{s}_1^\star), \ldots, Z(\bm{s}_m^\star) \} ^\top$. The kriging prediction, which is treated as an approximation of $Z(\bm{s})$, is 
\begin{equation*}
\text{E} \{ Z(\bm{s})|Z(\bm{s}_1^\star), \ldots, Z(\bm{s}_m^\star) \} = \bm{c}^\top(\bm{s}, \bm{\theta}) (C^\star)^{-1}(\bm{\theta}) \bm{Z}^\star, 
\end{equation*}
so the Gaussian predictive process model is 
\begin{equation} \label{Eq:GPP_model}
\tilde{Z}(\bm{s}) = \bm{c}^\top(\bm{s}, \bm{\theta}) (C^\star)^{-1}(\bm{\theta}) \bm{Z}^\star + \epsilon(\bm{s}), 
\end{equation}
where $\bm{c}(\bm{s}, \bm{\theta}) = [C(\bm{s}, \bm{s}_j^\star; \bm{\theta})]_{j=1}^m$, $C^\star(\bm{\theta})=[C(\bm{s}_i^\star, \bm{s}_j^\star; \bm{\theta})]_{i, j=1}^m$, $C(\bm{s}_1, \bm{s}_2; \bm{\theta}) = \text{Cov}\{Z(\bm{s}_1), Z(\bm{s}_2)\}$, and $\epsilon(\bm{s})$ is the nugget effect term which has a normal distribution with mean zero and variance $\tau^2$. The approximated kriging prediction is computed with the covariance matrix of the approximated random field $\tilde{Z}(\bm{s})$. For the Gaussian predictive process model, the computation of the inverse covariance matrix only involves the inversion of the matrix of order $m$, so computational time can be saved, compared with directly inverting a matrix of order $n$. 
In our comparison, we first fit the Gaussian predictive process model \eqref{Eq:GPP_model} by maximum likelihood estimation, where the covariance function $C(h; \bm{\theta})$ is the Mat\'{e}rn covariance \eqref{Eq:Matern_Cov}. The smoothness parameter $\nu$ is either treated as unknown, or fixed at $\nu=0.5$. Next, we compute the plug-in kriging prediction and the MSPE based on the GPP model,  on the same prediction locations used in the computation of Table \ref{Table:soil_result}. We choose the knots as the $23\times 23$ regular grid, which is evenly distributed on the observed range $[33.0, 35.2] \times [-106.1, -103.9]$. Results show that the MSPE of the GPP model is significantly larger than the corresponding prediction results based on exact MLE and has no apparent change when the number of observation $n$ increases. For instance, when $\nu$ is unknown and $n=3,600$, $14,400$, $32,400$, and $57,600$, the MSPE of the GPP model are $0.4292$, $0.4328$, $0.4298$, and $0.4268$, respectively, whereas the corresponding MSPE for the exact model are $0.2283$, $0.1461$, $0.1067$, and $0.0811$, respectively. 
We also consider the case with a larger number of knots $m$ for $n=3,600$. 
The knots used in this computation are chosen as the $\sqrt{m}\times \sqrt{m}$ regular grids, evenly distributed on the observed range $[33.0, 35.2] \times [-106.1, -103.9]$. Results show that even when $m=3,600$, the MSPE of the GPP method ($0.2406$ when $\nu$ is unknown or fixed) is still larger than that of the exact MLE method ($0.2283$ when $\nu$ is unknown or fixed). Thus, for kriging prediction of our considered soil dataset, the Gaussian predictive process method is less efficient than the exact MLE and the TLR approximations. 

In conclusion, our suggested settings of the tuning parameters for TLR approximation, obtained by using the MLOE, MMOM, and RMOM criteria, can maintain the estimation or prediction performances for the soil moisture data. 
Thus, we have successfully applied our suggested criteria to the TLR tuning parameter selection problem in applications. 
According to our comparison, the TLR  approximation with our suggested parameters outperforms the Gaussian predictive process method in the soil dataset prediction problem. 
For this soil data, the TLR approximation also outperforms the composite likelihood in approximating the exact MLE, though the MSPE of the TLR is here slightly less competitive. 
\section{Concluding Remarks} \label{sec:discussion}

In this article, we presented the Mean Loss of Efficiency (MLOE), Mean Misspecification of the MSE (MMOM), and Root mean square MOM (RMOM) criteria
as tools to detect the difference of the prediction performance between the true and the approximated covariance models in simulation studies. 
We found that the suggested criteria are more appropriate than the commonly used Mean Square Prediction Error criterion, 
as the criteria can detect the efficiency loss when a smoother covariance model is misspecified as a rougher covariance model with a nugget effect in simulation studies, 
which the MSPE cannot do.  
Our suggested criteria are valuable tools for understanding the impact of the tuning parameters on the statistical performance of sophisticated approximation methods, which is crucial for selecting these inputs. 
To illustrate this, 
we compared the estimation and prediction performances of the Tile Low-Rank (TLR) approximation with different tuning parameters, 
and obtained a practical suggestion on how to choose these tuning parameters for different application requirements. 
We showed by a real-case study in which our suggested tuning parameters obtained by our criteria works well to keep the estimation or prediction performances of the TLR method, e.g., the TLR outperforms the typical Gaussian predictive process method in prediction efficiency, and outperforms the composite likelihood in estimation efficiency.

It is worth noting that, the smoothness can affect the effectiveness of adopting the TLR approximation in spatial prediction and the proper value of tuning parameters in this approximation. 
For instance, if we can ensure that the process is not smoother than the Whittle covariance model, 
then we can set $tlr\_acc=10^{-7}$; else we may need a lower value, say $tlr\_acc=10^{-9}$, as is discussed in Remark \ref{Remark:smoother}. 
Thus, it would be appealing to introduce a suitable method for determining this kind of smoothness, such as determining the range of $\nu$ in the Mat\'{e}rn covariance model, before estimation and prediction. 
However, as it has been shown by simulations, misspecification of the smoothness of a random field significantly worsens the spatial prediction performance. 
Thus the smoothness determination method should be accurate enough. 
One can use a rough estimate of the smoothness, e.g., the composite likelihood method, for determining the tuning parameters. However, some methods, such as the TLR, require a range for unknown parameters as the input, rather than an initial value. In this case, determining a range of the smoothness parameter is more favorable. 
In future work, we will develop suitable smoothness parameter determination methods, 
such as the hypothesis tests proposed by \cite{hong2020hypothesis} and the references therein, 
and apply the method in the parameter estimation process to further improve the computation performance. 

Currently, the \exageostatr framework for computing the TLR method is not available for estimating the unknown nugget effect. In future work, we will try to overcome this restriction. After that, we will investigate the tuning parameter selections for this case. The fitting result of the soil moisture dataset may be better when a nugget effect term is involved in the spatial model. 
It would also be interesting to compare the performance of the other tile-based approximation methods with the TLR method, 
using our suggested criteria, 
in order to determine the best method for different application cases. 

\section{Acknowledgements}

The authors wish to thank the anonymous reviewers for their insightful comments and suggestions that substantially improved this paper.

This work was supported by the King Abdullah University of Science and Technology (KAUST) and partially supported by the NSFC (No. 11771241 and 11931001). 
\bibliographystyle{chicago}
\bibliography{mybibfile}

\newpage
\section{Supplementary Material}

In this Supplementary Material, we list the simulation results omitted in Section \ref{sec:experimental_results} due to space limitations. 

\begin{table}[htbp]
  \centering 
  \caption{Estimation and prediction performances of MLE and TLR approximation estimates for different $nb$ values, where the MLOE, MMOM, and RMOM criteria are computed by the plug-in method.  $\text{Bias}(\cdot)$ means the estimate of the parameter minus its true value, while the estimation time means the computational time of the corresponding estimation. The value of MLOE for all cases ($\nu=0.5$ and $\nu=1.0$) is multiplied by $10^6$. }  
  \resizebox{0.72\textwidth}{!}{
  \begin{tabular}{|c|c|c|cccc|c|cccc|}
  \hline 
  \multirow{3}{*}{$h_{\text{eff}}$} & \multirow{3}{*}{Mean (sd)} & \multicolumn{5}{c|}{$\nu=0.5$} & \multicolumn{5}{c|}{$\nu=1.0$} \\
    \cline{3-12}
      &    & \multirow{2}{*}{MLE} & \multicolumn{4}{c|}{Tile size ($nb$)} & \multirow{2}{*}{MLE} & \multicolumn{4}{c|}{Tile size ($nb$)} \\
  \cline{4-7} 
  \cline{9-12} 
  	&	&	& 400 & 450 & 600 & 900 &  & 400 & 450 & 600 & 900 \\
  \hline 
  \multirow{10}{*}{$0.2$} & \multirow{2}{*}{$\text{Bias} (\sigma^2)$} & $-$0.0080 & $-$0.0079 & $-$0.0079 & $-$0.0079 & $-$0.0079 & $-$0.0085 & $-$0.0061 & $-$0.0061 & $-$0.0061 & $-$0.0063 \\ 
      &    & (0.0908) & (0.0908) & (0.0908) & (0.0908) & (0.0908) & (0.1054) & (0.1058) & (0.1058) & (0.1058) & (0.1057) \\ 
      \cline{3-12}
      & \multirow{2}{*}{$\text{Bias} (\alpha)$} & $-$0.0006 & $-$0.0006 & $-$0.0006 & $-$0.0006 & $-$0.0006 & $-$0.0003 & $-$0.0002 & $-$0.0002 & $-$0.0002 & $-$0.0002 \\ 
      &    & (0.0063) & (0.0063) & (0.0063) & (0.0063) & (0.0063) & (0.0028) & (0.0029) & (0.0028) & (0.0029) & (0.0028) \\ 
      \cline{2-12}
      & \multirow{2}{*}{MLOE $(\times 10^6)$} & 3.3945 & 3.3756 & 3.3756 & 3.3758 & 3.3756 & 1.7378 & 1.6940 & 1.6938 & 1.6941 & 1.6915 \\ 
      &    & (5.9930) & (5.9474) & (5.9474) & (5.9477) & (5.9475) & (2.7659) & (2.6485) & (2.6479) & (2.6485) & (2.6499) \\ 
      \cline{3-12}
      & \multirow{2}{*}{MMOM} & 0.0017 & 0.0011 & 0.0011 & 0.0011 & 0.0011 & 0.0019 & $-$0.0016 & $-$0.0016 & $-$0.0016 & $-$0.0016 \\ 
      &    & (0.0232) & (0.0232) & (0.0232) & (0.0232) & (0.0232) & (0.0240) & (0.0240) & (0.0240) & (0.0240) & (0.0240) \\ 
      \cline{3-12}
      & \multirow{2}{*}{RMOM} & 0.0185 & 0.0185 & 0.0185 & 0.0185 & 0.0185 & 0.0189 & 0.0190 & 0.0190 & 0.0190 & 0.0190 \\ 
      &    & (0.0141) & (0.0140) & (0.0140) & (0.0140) & (0.0140) & (0.0147) & (0.0146) & (0.0146) & (0.0146) & (0.0146) \\ 
      \cline{2-12}
      & Estimation & 146.5 & 110.2 & 90.3 & 146.4 & 146.6 & 197.4 & 119.5 & 117.2 & 174.1 & 165.7 \\ 
      & time (sec) & (20.2) & (15.3) & (13.5) & (21.7) & (19.5) & (30.8) & (17.6) & (20.0) & (26.7) & (25.9) \\ 
  \hline 
  \multirow{10}{*}{$0.4$} & \multirow{2}{*}{$\text{Bias} (\sigma^2)$} & $-$0.0178 & $-$0.0172 & $-$0.0172 & $-$0.0172 & $-$0.0172 & $-$0.0207 & $-$0.0070 & $-$0.0070 & $-$0.0070 & $-$0.0070 \\ 
      &    & (0.1739) & (0.1741) & (0.1742) & (0.1742) & (0.1741) & (0.1955) & (0.1990) & (0.1990) & (0.1990) & (0.1990) \\ 
      \cline{3-12}
      & \multirow{2}{*}{$\text{Bias} (\alpha)$} & $-$0.0026 & $-$0.0023 & $-$0.0023 & $-$0.0023 & $-$0.0023 & $-$0.0016 & $-$0.0003 & $-$0.0003 & $-$0.0003 & $-$0.0003 \\ 
      &    & (0.0234) & (0.0234) & (0.0234) & (0.0234) & (0.0234) & (0.0100) & (0.0102) & (0.0102) & (0.0102) & (0.0102) \\ 
      \cline{2-12}
      & \multirow{2}{*}{MLOE $(\times 10^6)$} & 0.9790 & 0.9681 & 0.9682 & 0.9682 & 0.9681 & 0.4034 & 0.3659 & 0.3659 & 0.3659 & 0.3659 \\
      &    & (2.0811) & (2.0581) & (2.0581) & (2.0581) & (2.0581) & (0.6804) & (0.5944) & (0.5943) & (0.5943) & (0.5943) \\ 
      \cline{3-12}
      & \multirow{2}{*}{MMOM} & 0.0018 & 0.0005 & 0.0005 & 0.0005 & 0.0005 & 0.0019 & $-$0.0107 & $-$0.0107 & $-$0.0107 & $-$0.0107 \\ 
      &    & (0.0228) & (0.0228) & (0.0228) & (0.0228) & (0.0228) & (0.0230) & (0.0230) & (0.0230) & (0.0230) & (0.0230) \\ 
      \cline{3-12}
      & \multirow{2}{*}{RMOM} & 0.0183 & 0.0182 & 0.0182 & 0.0182 & 0.0182 & 0.0185 & 0.0206 & 0.0206 & 0.0206 & 0.0206 \\ 
      &    & (0.0136) & (0.0135) & (0.0135) & (0.0135) & (0.0135) & (0.0137) & (0.0147) & (0.0147) & (0.0147) & (0.0147) \\ 
      \cline{2-12}
      & Estimation & 115.8 & 83.6 & 72.6 & 114.5 & 116.3 & 228.4 & 115.7 & 107.7 & 156.5 & 173.0 \\ 
      & time (sec) & (18.0) & (14.2) & (14.2) & (22.0) & (22.9) & (43.0) & (23.8) & (26.0) & (32.2) & (41.5) \\ 
  \hline 
  \multirow{10}{*}{$0.8$} & \multirow{2}{*}{$\text{Bias} (\sigma^2)$} & $-$0.0147 & $-$0.0133 & $-$0.0133 & $-$0.0133 & $-$0.0133 & $-$0.0226 & 0.0384 & 0.0379 & 0.0384 & 0.0403 \\ 
      &    & (0.3474) & (0.3482) & (0.3482) & (0.3482) & (0.3482) & (0.3678) & (0.3925) & (0.3923) & (0.3927) & (0.3931) \\ 
      \cline{3-12}
      & \multirow{2}{*}{$\text{Bias} (\alpha)$} & $-$0.0042 & $-$0.0032 & $-$0.0032 & $-$0.0032 & $-$0.0032 & $-$0.0057 & 0.0050 & 0.0050 & 0.0050 & 0.0052 \\ 
      &    & (0.0928) & (0.0933) & (0.0933) & (0.0933) & (0.0933) & (0.0361) & (0.0383) & (0.0383) & (0.0383) & (0.0383) \\ 
      \cline{2-12}
      & \multirow{2}{*}{MLOE $(\times 10^6)$} & 0.3286 & 0.3221 & 0.3221 & 0.3221 & 0.3220 & 0.1051 & 0.0759 & 0.0758 & 0.0758 & 0.0759 \\ 
      &    & (0.9662) & (0.9494) & (0.9494) & (0.9494) & (0.9494) & (0.2078) & (0.1429) & (0.1427) & (0.1427) & (0.1426) \\ 
      \cline{3-12}
      & \multirow{2}{*}{MMOM} & 0.0016 & $-$0.0007 & $-$0.0007 & $-$0.0007 & $-$0.0007 & 0.0017 & $-$0.0438 & $-$0.0438 & $-$0.0438 & $-$0.0438 \\ 
      &    & (0.0227) & (0.0227) & (0.0227) & (0.0227) & (0.0227) & (0.0225) & (0.0225) & (0.0225) & (0.0225) & (0.0225) \\ 
      \cline{3-12}
      & \multirow{2}{*}{RMOM} & 0.0182 & 0.0182 & 0.0182 & 0.0182 & 0.0182 & 0.0182 & 0.0446 & 0.0446 & 0.0446 & 0.0446 \\ 
      &    & (0.0135) & (0.0135) & (0.0135) & (0.0135) & (0.0135) & (0.0135) & (0.0210) & (0.0210) & (0.0210) & (0.0210) \\ 
      \cline{2-12}
      & Estimation & 145.8 & 110.5 & 96.1 & 142.5 & 147.6 & 218.4 & 144.2 & 120.1 & 167.3 & 211.8 \\ 
      & time (sec) & (26.4) & (19.4) & (16.4) & (22.0) & (25.2) & (51.9) & (36.7) & (33.0) & (40.9) & (53.2) \\ 
  \hline 
  \multirow{10}{*}{$1.6$} & \multirow{2}{*}{$\text{Bias} (\sigma^2)$} & 0.0132 & 0.0158 & 0.0158 & 0.0159 & 0.0156 & 0.0163 & 0.2173 & 0.2236 & 0.2210 & 0.2153 \\ 
      &    & (0.6292) & (0.6308) & (0.6309) & (0.6309) & (0.6300) & (0.6546) & (0.7339) & (0.7793) & (0.7624) & (0.7419) \\ 
      \cline{3-12}
      & \multirow{2}{*}{$\text{Bias} (\alpha)$} & 0.0069 & 0.0108 & 0.0108 & 0.0108 & 0.0107 & $-$0.0144 & 0.0577 & 0.0577 & 0.0579 & 0.0571 \\ 
      &    & (0.3369) & (0.3393) & (0.3393) & (0.3393) & (0.3388) & (0.1186) & (0.1348) & (0.1394) & (0.1370) & (0.1357) \\ 
      \cline{2-12}
      & \multirow{2}{*}{MLOE $(\times 10^6)$} & 0.1221 & 0.1178 & 0.1178 & 0.1178 & 0.1178 & 0.0273 & 0.0109 & 0.0110 & 0.0109 & 0.0109 \\ 
      &    & (0.4450) & (0.4307) & (0.4307) & (0.4307) & (0.4307) & (0.0669) & (0.0242) & (0.0243) & (0.0242) & (0.0241) \\ 
      \cline{3-12}
      & \multirow{2}{*}{MMOM} & 0.0014 & $-$0.0033 & $-$0.0033 & $-$0.0033 & $-$0.0033 & 0.0014 & $-$0.1428 & $-$0.1428 & $-$0.1428 & $-$0.1428 \\ 
      &    & (0.0227) & (0.0227) & (0.0227) & (0.0227) & (0.0227) & (0.0227) & (0.0223) & (0.0222) & (0.0223) & (0.0223) \\ 
      \cline{3-12}
      & \multirow{2}{*}{RMOM} & 0.0181 & 0.0182 & 0.0182 & 0.0182 & 0.0182 & 0.0182 & 0.1428 & 0.1428 & 0.1428 & 0.1428 \\ 
      &    & (0.0136) & (0.0138) & (0.0138) & (0.0138) & (0.0138) & (0.0135) & (0.0223) & (0.0222) & (0.0223) & (0.0223) \\ 
      \cline{2-12}
      & Estimation & 161.9 & 111.4 & 95.8 & 145.5 & 161.5 & 277.5 & 122.3 & 108.1 & 143.5 & 186.5 \\ 
      & time (sec) & (28.2) & (22.8) & (20.8) & (28.8) & (32.0) & (75.6) & (35.6) & (31.6) & (46.6) & (63.0) \\ 
  \hline 
  \end{tabular}}
  \label{Table:nb_estimate_full} 
\end{table}

\begin{table}[htbp]
  \centering 
  \caption{Estimation and prediction performances of the exact MLE and TLR approximation estimates for different $tlr\_acc$ values, where the MLOE, MMOM, and RMOM criteria are computed by the plug-in method.  $\text{Bias}(\cdot)$ means the estimate of the parameter minus its true value, while the estimation time means the computational time of the corresponding estimation. The value of MLOE for all cases ($\nu=0.5$ and $\nu=1.0$) is multiplied by $10^6$. The missing part in the table (-) means that the result is not available, because the covariance matrix is numerically non positive-definite. } 
  \resizebox{0.95\textwidth}{!}{
  \begin{tabular}{|c|c|c|cccc|c|cccc|}
  \hline
  \multirow{3}{*}{$h_{\text{eff}}$} & \multirow{3}{*}{Mean (sd)} & \multicolumn{5}{c|}{$\nu=0.5$} & \multicolumn{5}{c|}{$\nu=1.0$} \\
    \cline{3-12}
      &    & \multirow{2}{*}{MLE} & \multicolumn{4}{c|}{TLR accuracy ($tlr\_acc$)} & \multirow{2}{*}{MLE} & \multicolumn{4}{c|}{TLR accuracy ($tlr\_acc$)} \\
  \cline{4-7} 
  \cline{9-12} 
  	&	&	& $10^{-5}$ & $10^{-7}$ & $10^{-9}$ & $10^{-11}$ &  & $10^{-5}$ & $10^{-7}$ & $10^{-9}$ & $10^{-11}$ \\
  \hline
  \multirow{12}{*}{$0.2$} & \multirow{2}{*}{$\text{Bias} (\sigma^2)$} & $-$0.0080 & $-$0.0023 & $-$0.0079 & $-$0.0079 & $-$0.0079 & $-$0.0085 & 0.0124 & $-$0.0061 & $-$0.0061 & $-$0.0061 \\ 
  									  & 											  & (0.0908) & (0.1095) & (0.0908) & (0.0908) & (0.0908) & (0.1054) & (0.1294) & (0.1058) & (0.1058) & (0.1058) \\ 
  \cline{3-12}
  									  & \multirow{2}{*}{$\text{Bias} (\alpha)$} & $-$0.0006 & $-$0.0002 & $-$0.0006 & $-$0.0006 & $-$0.0006 & $-$0.0003 & 0.0003 & $-$0.0002 & $-$0.0002 & $-$0.0002 \\ 
  									  & 										  & (0.0063) & (0.0077) & (0.0063) & (0.0063) & (0.0063) & (0.0028) & (0.0034) & (0.0029) & (0.0028) & (0.0029) \\ 
  \cline{2-12}
      & \multirow{2}{*}{MLOE $(\times 10^6)$} & 3.3945 & 3.5691 & 3.3756 & 3.3756 & 3.3757 &  1.7378 & 2.1075 & 1.6942 & 1.6938 & 1.6940 \\ 
      &    & (5.9931) & (6.4517) & (5.9476) & (5.9474) & (5.9474) & (2.7659) & (2.9479) & (2.6494) & (2.6479) & (2.6485) \\ 
  \cline{3-12}
      & \multirow{2}{*}{MMOM} & 0.0017 & 0.0009 & 0.0011 & 0.0011 & 0.0011 & 0.0019 & $-$0.0022 & $-$0.0016 & $-$0.0016 & $-$0.0016 \\ 
      &    & (0.0232) & (0.0233) & (0.0232) & (0.0232) & (0.0232) & (0.0240) & (0.0243) & (0.0240) & (0.0240) & (0.0240) \\ 
  \cline{3-12}
      & \multirow{2}{*}{RMOM} & 0.0185 & 0.0186 & 0.0185 & 0.0185 & 0.0185 & 0.0189 & 0.0194 & 0.0190 & 0.0190 & 0.0190 \\ 
      &    & (0.0141) & (0.0140) & (0.0140) & (0.0140) & (0.0140) & (0.0147) & (0.0146) & (0.0146) & (0.0146) & (0.0146) \\ 
  \cline{2-12}
  									  & Estimation & 168.1& 69.8 & 77.8 & 90.4 & 112.1 & 211.0 & 102.9 & 108.5 & 116.3 & 125.3 \\ 
  									  & time (sec) & (22.3) & (11.9) & (9.5) & (13.5) & (15.5) & (34.0) & (25.2) & (16.9) & (19.8) & (22.0) \\ 
  \hline
  \multirow{12}{*}{$0.4$} & \multirow{2}{*}{$\text{Bias} (\sigma^2)$} & $-$0.0178 & $-$0.0073 & $-$0.0172 & $-$0.0172 & $-$0.0172 & $-$0.0207 & - & $-$0.0047 & $-$0.0070 & $-$0.0070 \\ 
    									  & 											  & (0.1739) & (0.1668) & (0.1741) & (0.1742) & (0.1741) & (0.1955) & - & (0.1992) & (0.1990) & (0.1990) \\ 
	 \cline{3-12}
    									& \multirow{2}{*}{$\text{Bias} (\alpha)$} & $-$0.0026 & $-$0.0010 & $-$0.0023 & $-$0.0023 & $-$0.0023 & $-$0.0016 & - & $-$0.0002 & $-$0.0003 & $-$0.0003 \\ 
  									& 										    & (0.0234) & (0.0224) & (0.0234) & (0.0234) & (0.0234) & (0.0100) & - & (0.0102) & (0.0102) & (0.0102) \\ 
	 \cline{2-12}
      & \multirow{2}{*}{MLOE $(\times 10^6)$} & 0.9790 & 0.9290 & 0.9679 & 0.9682 & 0.9681 & 0.4034 & - & 0.3665 & 0.3659 & 0.3659 \\ 
      &    & (2.0811) & (2.0983) & (2.0559) & (2.0581) & (2.0581) & (0.6804) & - & (0.5937) & (0.5943) & (0.5943) \\  
  \cline{3-12}
      & \multirow{2}{*}{MMOM} & 0.0018 & 0.0004 & 0.0005 & 0.0005 & 0.0005 & 0.0019 & - & $-$0.0107 & $-$0.0107 & $-$0.0107 \\ 
      &    & (0.0228) & (0.0227) & (0.0228) & (0.0228) & (0.0228) & (0.0231) & - & (0.0230) & (0.0230) & (0.0230) \\ 
  \cline{3-12}
      & \multirow{2}{*}{RMOM} & 0.0183 & 0.0182 & 0.0182 & 0.0182 & 0.0182 & 0.0185 & - & 0.0206 & 0.0206 & 0.0206 \\ 
      &    & (0.0136) & (0.0135) & (0.0135) & (0.0135) & (0.0135) & (0.0137) & - & (0.0147) & (0.0147) & (0.0147) \\ 
  \cline{2-12}
  									& Estimation & 126.8 & 60.2 & 65.2 & 72.4 & 86.9 & 252.1 & - & 103.0 & 107.5 & 117.2 \\ 
  									& time (sec) & (19.5) & (15.6) & (12.9) & (14.1) & (14.7) & (49.2) & - & (23.6) & (26.0) & (24.1) \\ 
  \hline
  \multirow{12}{*}{$0.8$} & \multirow{2}{*}{$\text{Bias} (\sigma^2)$} & $-$0.0147 & 0.3231 & $-$0.0171 & $-$0.0133 & $-$0.0133 & $-$0.0226 & - & 0.0458 & 0.0379 & 0.0387 \\ 
    									  & 											  & (0.3474) & (0.2532) & (0.3368) & (0.3482) & (0.3482) & (0.3678) & - & (0.3241) & (0.3923) & (0.3934) \\ 
	 \cline{3-12}
    									  & \multirow{2}{*}{$\text{Bias} (\alpha)$} & $-$0.0042 & 0.0880 & $-$0.0043 & $-$0.0032 & $-$0.0032 & $-$0.0057 & - & 0.0068 & 0.0050 & 0.0050 \\ 
    									  & 										  & (0.0928) & (0.0704) & (0.0903) & (0.0933) & (0.0933) & (0.0361) & - & (0.0328) & (0.0383) & (0.0383) \\ 
	 \cline{2-12}
      & \multirow{2}{*}{MLOE $(\times 10^6)$} & 0.3286 & 0.1459 & 0.3213 & 0.3221 & 0.3220 & 0.1051 & - & 0.0624 & 0.0758 & 0.0759 \\ 
      &    & (0.9662) & (0.2913) & (0.9496) & (0.9494) & (0.9494) & (0.2078) & - & (0.1390) & (0.1427) & (0.1427) \\ 
  \cline{3-12}
      & \multirow{2}{*}{MMOM} & 0.0016 & $-$0.0028 & $-$0.0007 & $-$0.0007 & $-$0.0007 & 0.0017 & - & $-$0.0441 & $-$0.0438 & $-$0.0438 \\ 
      &    & (0.0227) & (0.0230) & (0.0227) & (0.0227) & (0.0227) & (0.0227) & - & (0.0227) & (0.0225) & (0.0225) \\ 
  \cline{3-12}
      & \multirow{2}{*}{RMOM} & 0.0182 & 0.0185 & 0.0182 & 0.0182 & 0.0182 & 0.0182 & - & 0.0448 & 0.0446 & 0.0446 \\ 
      &    & (0.0135) & (0.0139) & (0.0135) & (0.0135) & (0.0135) & (0.0135) & - & (0.0211) & (0.0210) & (0.0210) \\ 
  \cline{2-12}
    									  & Estimation & 159.3 & 70.1 & 86.5 & 96.2 & 112.9 & 235.5 & - & 106.1 & 120.4 & 130.9 \\ 
    									  & time (sec) & (28.4) & (25.9) & (17.7) & (16.4) & (18.6) & (54.1) & - & (36.6) & (33.2) & (33.2) \\ 
  \hline
  \multirow{12}{*}{$1.6$} & \multirow{2}{*}{$\text{Bias} (\sigma^2)$} & 0.0133 & 0.1515 & $-$0.0066 & 0.0158 & 0.0159 & 0.0163 & - & 0.3800 & 0.2236 & 0.2276 \\  
    									  & 											  & (0.6292) & (0.1781) & (0.5604) & (0.6309) & (0.6309) & (0.6546) & - & (0.3154) & (0.7793) & (0.7927) \\ 
	 \cline{3-12}
    									  & \multirow{2}{*}{$\text{Bias} (\alpha)$} & 0.0069 & 0.0843 & $-$0.0013 & 0.0108 & 0.0108 & $-$0.0144 & - & 0.1033 & 0.0577 & 0.0581 \\ 
    									  & 										  & (0.3369) & (0.0993) & (0.3011) & (0.3393) & (0.3394) & (0.1186) & - & (0.0686) & (0.1394) & (0.1408) \\ 
	 \cline{2-12}
      & \multirow{2}{*}{MLOE $(\times 10^6)$} & 0.1221 & 0.0044 & 0.1174 & 0.1178 & 0.1178 & 0.0273 & - & 0.0079 & 0.0110 & 0.0109 \\ 
      &    & (0.4450) & (0.0223) & (0.4307) & (0.4307) & (0.4307) & (0.0669) & - & (0.0167) & (0.0243) & (0.0242) \\ 
  \cline{3-12}
      & \multirow{2}{*}{MMOM} & 0.0014 & $-$0.0044 & $-$0.0033 & $-$0.0033 & $-$0.0033 & 0.0014 & - & $-$0.1436 & $-$0.1428 & $-$0.1428 \\ 
      &    & (0.0227) & (0.0227) & (0.0227) & (0.0227) & (0.0227) & (0.0227) & - & (0.0222) & (0.0222) & (0.0222) \\ 
  \cline{3-12}
      & \multirow{2}{*}{RMOM} & 0.0181 & 0.0183 & 0.0182 & 0.0182 & 0.0182 & 0.0182 & - & 0.1436 & 0.1428 & 0.1428 \\ 
      &    & (0.0136) & (0.0139) & (0.0138) & (0.0138) & (0.0138) & (0.0135) & - & (0.0222) & (0.0222) & (0.0222) \\ 
  \cline{2-12}
    									  & Estimation & 177.5 & 68.7 & 93.8 & 96.3 & 117.1 & 274.3 & - & 62.7 & 106.7 & 111.6 \\ 
    									  & time (sec) & (33.9) & (20.6) & (18.5) & (20.9) & (22.4) & (74.6) & - & (26.7) & (31.2) & (33.7) \\ 
  \hline
  \end{tabular}}
  \label{Table:tlracc_estimate_full} 
\end{table}

\begin{table}[htbp]
  \centering 
  \caption{Estimation and prediction performances of the exact MLE and TLR approximation estimates for different $opt\_tol$ values, where the MLOE, MMOM, and RMOM criteria are computed by the plug-in method.  $\text{Bias}(\cdot)$ means the estimate of the parameter minus its true value, while the estimation time means the computational time of the corresponding estimation. The value of MLOE for all cases ($\nu=0.5$ and $\nu=1.0$) is multiplied by $10^6$. } 
  \resizebox{\textwidth}{!}{
  \begin{tabular}{|c|c|c|cccc|c|cccc|}
  \hline
  \multirow{3}{*}{$h_{\text{eff}}$} & \multirow{3}{*}{Mean (sd)} & \multicolumn{5}{c|}{$\nu=0.5$} & \multicolumn{5}{c|}{$\nu=1.0$} \\
  \cline{3-12}
      &    & \multirow{2}{*}{MLE} & \multicolumn{4}{c|}{Optimization tolerance ($opt\_tol$)} & \multirow{2}{*}{MLE} & \multicolumn{4}{c|}{Optimization tolerance ($opt\_tol$)} \\
  \cline{4-7} 
  \cline{9-12}
  	&	&	& $10^{-3}$ & $10^{-6}$ & $10^{-9}$ & $10^{-12}$ &	& $10^{-3}$ & $10^{-6}$ & $10^{-9}$ & $10^{-12}$ \\
  \hline
  \multirow{12}{*}{$0.2$} & \multirow{2}{*}{$\text{Bias} (\sigma^2)$} & $-$0.0080 & 0.3654 & $-$0.0079 & $-$0.0079 & $-$0.0079 & $-$0.0085 & 0.2977 & $-$0.0061 & $-$0.0061 & $-$0.0061 \\ 
      &    & (0.0908) & (0.3017) & (0.0908) & (0.0908) & (0.0908) & (0.1054) & (0.0732) & (0.1058) & (0.1058) & (0.1058) \\ 
       \cline{3-12} 
      & \multirow{2}{*}{$\text{Bias} (\alpha)$} & $-$0.0006 & 0.0253 & $-$0.0006 & $-$0.0006 & $-$0.0006 & $-$0.0003 & 0.0073 & $-$0.0002 & $-$0.0002 & $-$0.0002 \\ 
      &    & (0.0063) & (0.0210) & (0.0063) & (0.0063) & (0.0063) & (0.0028) & (0.0020) & (0.0028) & (0.0029) & (0.0029) \\ 
      \cline{2-12} 
      & \multirow{2}{*}{MLOE $(\times 10^6)$} & 3.3945 & 19.3107 & 3.3756 & 3.3756 & 3.3756 & 1.7378 & 6.9237 & 1.6938 & 1.6940 & 1.6940 \\ 
      &    & (5.9931) & (13.9768) & (5.9474) & (5.9475) & (5.9475) & (2.7659) & (2.9270) & (2.6479) & (2.6485) & (2.6485) \\ 
      \cline{3-12}
      & \multirow{2}{*}{MMOM} & 0.0017 & $-$0.0061 & 0.0011 & 0.0011 & 0.0011 & 0.0019 & $-$0.0103 & $-$0.0016 & $-$0.0016 & $-$0.0016 \\ 
      &    & (0.0232) & (0.0234) & (0.0232) & (0.0232) & (0.0232) & (0.0240) & (0.0243) & (0.0240) & (0.0240) & (0.0240) \\ 
      \cline{3-12}
      & \multirow{2}{*}{RMOM} & 0.0185 & 0.0196 & 0.0185 & 0.0185 & 0.0185 & 0.0189 & 0.0215 & 0.0190 & 0.0190 & 0.0190 \\ 
      &    & (0.0141) & (0.0140) & (0.0140) & (0.0140) & (0.0140) & (0.0147) & (0.0153) & (0.0146) & (0.0146) & (0.0146) \\ 
      \cline{2-12} 
      & Estimation & 168.1 & 33.8 & 90.4 & 102.2 & 113.0 & 211.0 & 20.7 & 116.3 & 133.3 & 146.9 \\ 
      & time (sec) & (22.3) & (13.6) & (13.5) & (13.2) & (13.0) & (34.0) & (6.5) & (19.8) & (20.3) & (20.2) \\ 
  \hline 
  \multirow{12}{*}{$0.4$} & \multirow{2}{*}{$\text{Bias} (\sigma^2)$} & $-$0.0178 & 0.0836 & $-$0.0172 & $-$0.0172 & $-$0.0172 & $-$0.0207 & 0.1373 & $-$0.0070 & $-$0.0070 & $-$0.0070 \\ 
      &    & (0.1739) & (0.0793) & (0.1742) & (0.1741) & (0.1741) & (0.1955) & (0.0731) & (0.1990) & (0.1990) & (0.1990) \\ 
       \cline{3-12} 
      & \multirow{2}{*}{$\text{Bias} (\alpha)$} & $-$0.0026 & 0.0114 & $-$0.0023 & $-$0.0023 & $-$0.0023 & $-$0.0016 & 0.0072 & $-$0.0003 & $-$0.0003 & $-$0.0003 \\ 
      &    & (0.0234) & (0.0111) & (0.0234) & (0.0234) & (0.0234) & (0.0100) & (0.0029) & (0.0102) & (0.0102) & (0.0102) \\ 
      \cline{2-12} 
      & \multirow{2}{*}{MLOE $(\times 10^6)$} & 0.9790 & 0.3301 & 0.9682 & 0.9681 & 0.9681 & 0.4034 & 0.1364 & 0.3659 & 0.3659 & 0.3659 \\ 
      &    & (2.0811) & (1.1812) & (2.0581) & (2.0581) & (2.0581) & (0.6804) & (0.1295) & (0.5943) & (0.5943) & (0.5943) \\ 
      \cline{3-12}
      & \multirow{2}{*}{MMOM} & 0.0018 & $-$0.0009 & 0.0005 & 0.0005 & 0.0005 & 0.0019 & $-$0.0105 & $-$0.0107 & $-$0.0107 & $-$0.0107 \\ 
      &    & (0.0228) & (0.0225) & (0.0228) & (0.0228) & (0.0228) & (0.0231) & (0.0222) & (0.0230) & (0.0230) & (0.0230) \\ 
      \cline{3-12}
      & \multirow{2}{*}{RMOM} & 0.0183 & 0.0179 & 0.0182 & 0.0182 & 0.0182 & 0.0185 & 0.0198 & 0.0206 & 0.0206 & 0.0206 \\ 
      &    & (0.0136) & (0.0135) & (0.0135) & (0.0135) & (0.0135) & (0.0137) & (0.0145) & (0.0147) & (0.0147) & (0.0147) \\ 
      \cline{2-12} 
      & Estimation & 126.8 & 19.8 & 72.4 & 85.6 & 97.8 & 252.1 & 22.4 & 107.5 & 126.4 & 142.2 \\ 
      & time (sec) & (19.5) & (5.5) & (14.1) & (14.0) & (14.2) & (49.2) & (2.8) & (26.0) & (24.5) & (24.9) \\ 
  \hline 
  \multirow{12}{*}{$0.8$} & \multirow{2}{*}{$\text{Bias} (\sigma^2)$} & $-$0.0147 & 0.2387 & $-$0.0133 & $-$0.0133 & $-$0.0133 & $-$0.0226 & 0.1099 & 0.0379 & 0.0381 & 0.0381 \\ 
      &    & (0.3474) & (0.3109) & (0.3482) & (0.3482) & (0.3482) & (0.3678) & (0.0449) & (0.3923) & (0.3925) & (0.3925) \\ 
       \cline{3-12} 
      & \multirow{2}{*}{$\text{Bias} (\alpha)$} & $-$0.0042 & 0.0649 & $-$0.0032 & $-$0.0032 & $-$0.0032 & $-$0.0057 & 0.0156 & 0.0050 & 0.0050 & 0.0050 \\ 
      &    & (0.0928) & (0.0853) & (0.0933) & (0.0933) & (0.0933) & (0.0361) & (0.0056) & (0.0383) & (0.0383) & (0.0383) \\ 
      \cline{2-12} 
      & \multirow{2}{*}{MLOE $(\times 10^6)$} & 0.3286 & 0.2252 & 0.3221 & 0.3221 & 0.3221 & 0.1051 & 0.0102 & 0.0758 & 0.0758 & 0.0758 \\ 
      &    & (0.9662) & (0.6756) & (0.9494) & (0.9494) & (0.9494) & (0.2078) & (0.0074) & (0.1427) & (0.1427) & (0.1427) \\ 
      \cline{3-12}
      & \multirow{2}{*}{MMOM} & 0.0016 & $-$0.0018 & $-$0.0007 & $-$0.0007 & $-$0.0007 & 0.0017 & $-$0.0448 & $-$0.0438 & $-$0.0438 & $-$0.0438 \\ 
      &    & (0.0227) & (0.0229) & (0.0227) & (0.0227) & (0.0227) & (0.0227) & (0.0226) & (0.0225) & (0.0225) & (0.0225) \\ 
      \cline{3-12}
      & \multirow{2}{*}{RMOM} & 0.0182 & 0.0182 & 0.0182 & 0.0182 & 0.0182 & 0.0182 & 0.0455 & 0.0446 & 0.0446 & 0.0446 \\ 
      &    & (0.0135) & (0.0140) & (0.0135) & (0.0135) & (0.0135) & (0.0135) & (0.0210) & (0.0210) & (0.0210) & (0.0210) \\ 
      \cline{2-12} 
      & Estimation & 159.3 & 32.3 & 96.2 & 108.6 & 121.3 & 235.5 & 21.4 & 120.4 & 137.1 & 152.3 \\ 
      & time (sec) & (28.4) & (16.9) & (16.4) & (17.3) & (17.1) & (54.1) & (6.4) & (33.2) & (32.6) & (32.5) \\ 
  \hline 
  \multirow{12}{*}{$1.6$} & \multirow{2}{*}{$\text{Bias} (\sigma^2)$} & 0.0133 & 0.0962 & 0.0158 & 0.0158 & 0.0158 & 0.0163 & 0.3263 & 0.2236 & 0.2234 & 0.2234 \\ 
      &    & (0.6292) & (0.4611) & (0.6309) & (0.6309) & (0.6309) & (0.6546) & (0.4421) & (0.7793) & (0.7787) & (0.7787) \\ 
       \cline{3-12} 
      & \multirow{2}{*}{$\text{Bias} (\alpha)$} & 0.0069 & 0.0542 & 0.0108 & 0.0108 & 0.0108 & $-$0.0144 & 0.0896 & 0.0577 & 0.0577 & 0.0577 \\ 
      &    & (0.3369) & (0.2483) & (0.3393) & (0.3393) & (0.3393) & (0.1186) & (0.0904) & (0.1394) & (0.1393) & (0.1393) \\ 
      \cline{2-12} 
      & \multirow{2}{*}{MLOE $(\times 10^6)$} & 0.1221 & 0.0694 & 0.1178 & 0.1178 & 0.1178 & 0.0273 & 0.0150 & 0.0110 & 0.0110 & 0.0110 \\ 
      &    & (0.4450) & (0.3458) & (0.4307) & (0.4307) & (0.4307) & (0.0669) & (0.0708) & (0.0243) & (0.0243) & (0.0243) \\ 
      \cline{3-12}
      & \multirow{2}{*}{MMOM} & 0.0014 & $-$0.0038 & $-$0.0033 & $-$0.0033 & $-$0.0033 & 0.0014 & $-$0.1432 & $-$0.1428 & $-$0.1428 & $-$0.1428 \\ 
      &    & (0.0227) & (0.0227) & (0.0227) & (0.0227) & (0.0227) & (0.0227) & (0.0220) & (0.0222) & (0.0222) & (0.0222) \\ 
      \cline{3-12}
      & \multirow{2}{*}{RMOM} & 0.0181 & 0.0184 & 0.0182 & 0.0182 & 0.0182 & 0.0182 & 0.1432 & 0.1428 & 0.1428 & 0.1428 \\ 
      &    & (0.0136) & (0.0137) & (0.0138) & (0.0138) & (0.0138) & (0.0135) & (0.0220) & (0.0222) & (0.0222) & (0.0222) \\ 
      \cline{2-12} 
      & Estimation & 177.5 & 42.0 & 96.3 & 113.2 & 126.8 & 274.3 & 29.4 & 106.7 & 124.1 & 136.2 \\ 
      & time (sec) & (33.9) & (18.5) & (20.9) & (21.4) & (21.3) & (74.6) & (21.7) & (31.2) & (34.8) & (36.0) \\ 
  \hline 
  \end{tabular}}
  \label{Table:prectlr_estimate_full} 
\end{table}

\begin{table}[htbp]
  \centering
  \caption{Prediction performance and the computational time for Tile Low-Rank approximations with different combinations of $tlr\_acc$ and $opt\_tol$, where $\nu$ is the smoothness parameter and $h_{\text{eff}}$ is the effective range. The MLOE, MMOM, and RMOM criteria are computed by the plug-in method. The estimation time means the computational time of the corresponding estimation, while value of MLOE for all cases ($\nu=0.5$ and $\nu=1.0$) is multiplied by $10^6$. } 
  \resizebox{\textwidth}{!}{
  \begin{tabular}{|c|c|cccc|cccc|}
  \hline 
  \multirow{2}{*}{$h_{\text{eff}}$} & \multirow{2}{*}{Mean (sd)} & \multicolumn{4}{c|}{$(tlr\_acc, opt\_tol)$, $\nu=0.5$} & \multicolumn{4}{c|}{$(tlr\_acc, opt\_tol)$, $\nu=1.0$} \\
  \cline{3-10}
      &    & $(10^{-7}, 10^{-3})$ & $(10^{-9}, 10^{-3})$ & $(10^{-7}, 10^{-6})$ & $(10^{-9}, 10^{-6})$ & $(10^{-7}, 10^{-3})$ & $(10^{-9}, 10^{-3})$ & $(10^{-7}, 10^{-6})$ & $(10^{-9}, 10^{-6})$ \\
  \hline 
  \multirow{8}{*}{$0.2$} & \multirow{2}{*}{MLOE $(\times 10^6)$} & 19.0060 & 19.3107 & 3.3756 & 3.3756 & 7.0268 & 6.9237 & 1.6942 & 1.6938 \\ 
      &    & (13.6484) & (13.9768) & (5.9476) & (5.9474) & (2.8741) & (2.9270) & (2.6494) & (2.6479) \\ 
      \cline{3-10}
      & \multirow{2}{*}{MMOM} & $-$0.0062 & $-$0.0061 & 0.0011 & 0.0011 & $-$0.0106 & $-$0.0103 & $-$0.0016 & $-$0.0016 \\ 
      &    & (0.0232) & (0.0234) & (0.0232) & (0.0232) & (0.0246) & (0.0243) & (0.0240) & (0.0240) \\ 
      \cline{3-10}
      & \multirow{2}{*}{RMOM} & 0.0193 & 0.0196 & 0.0185 & 0.0185 & 0.0218 & 0.0215 & 0.0190 & 0.0190 \\ 
      &    & (0.0142) & (0.0140) & (0.0140) & (0.0140) & (0.0154) & (0.0153) & (0.0146) & (0.0146) \\ 
      \cline{2-10}
      & Estimation & 28.8 & 33.1 & 76.6 & 88.5 & 19.7 & 20.8 & 108.9 & 117.2 \\ 
      & time (sec) & (11.4) & (13.3) & (9.4) & (13.2) & (7.0) & (6.5) & (16.9) & (20.0) \\ 
  \hline 
  \multirow{8}{*}{$0.4$} & \multirow{2}{*}{MLOE $(\times 10^6)$} & 0.3374 & 0.3301 & 0.9679 & 0.9682 & 0.1348 & 0.1364 & 0.3665 & 0.3659 \\ 
      &    & (1.2055) & (1.1812) & (2.0559) & (2.0581) & (0.1289) & (0.1295) & (0.5937) & (0.5943) \\ 
      \cline{3-10}
      & \multirow{2}{*}{MMOM} & $-$0.0011 & $-$0.0009 & 0.0005 & 0.0005 & $-$0.0105 & $-$0.0105 & $-$0.0107 & $-$0.0107 \\ 
      &    & (0.0226) & (0.0225) & (0.0228) & (0.0228) & (0.0221) & (0.0222) & (0.0230) & (0.0230) \\ 
      \cline{3-10}
      & \multirow{2}{*}{RMOM} & 0.0180 & 0.0179 & 0.0182 & 0.0182 & 0.0198 & 0.0198 & 0.0206 & 0.0206 \\ 
      &    & (0.0136) & (0.0135) & (0.0135) & (0.0135) & (0.0143) & (0.0145) & (0.0147) & (0.0147) \\ 
      \cline{2-10}
      & Estimation & 17.8 & 19.7 & 65.1 & 72.6 & 21.0 & 22.5 & 103.3 & 108.1 \\ 
      & time (sec) & (5.2) & (5.5) & (13.0) & (14.2) & (1.6) & (2.8) & (23.6) & (26.1) \\ 
  \hline 
  \multirow{8}{*}{$0.8$} & \multirow{2}{*}{MLOE $(\times 10^6)$} & 0.1703 & 0.2252 & 0.3213 & 0.3221 & 0.0102 & 0.0102 & 0.0624 & 0.0758 \\ 
      &    & (0.3915) & (0.6756) & (0.9496) & (0.9494) & (0.0070) & (0.0074) & (0.1390) & (0.1427) \\ 
      \cline{3-10}
      & \multirow{2}{*}{MMOM} & $-$0.0019 & $-$0.0018 & $-$0.0007 & $-$0.0007 & $-$0.0447 & $-$0.0448 & $-$0.0441 & $-$0.0438 \\ 
      &    & (0.0230) & (0.0229) & (0.0227) & (0.0227) & (0.0227) & (0.0226) & (0.0227) & (0.0225) \\ 
      \cline{3-10}
      & \multirow{2}{*}{RMOM} & 0.0183 & 0.0182 & 0.0182 & 0.0182 & 0.0455 & 0.0455 & 0.0448 & 0.0446 \\ 
      &    & (0.0140) & (0.0140) & (0.0135) & (0.0135) & (0.0211) & (0.0210) & (0.0211) & (0.0210) \\ 
      \cline{2-10}
      & Estimation & 28.4 & 32.4 & 86.8 & 96.5 & 20.0 & 21.4 & 106.3 & 120.4 \\ 
      & time (sec) & (15.0) & (16.9) & (17.8) & (16.5) & (5.3) & (6.4) & (36.8) & (33.1) \\ 
  \hline 
  \multirow{8}{*}{$1.6$} & \multirow{2}{*}{MLOE $(\times 10^6)$} & 0.0760 & 0.0694 & 0.1174 & 0.1178 & 0.0138 & 0.0150 & 0.0079 & 0.0110 \\ 
      &    & (0.3605) & (0.3458) & (0.4307) & (0.4307) & (0.0688) & (0.0708) & (0.0167) & (0.0243) \\ 
      \cline{3-10}
      & \multirow{2}{*}{MMOM} & $-$0.0038 & $-$0.0038 & $-$0.0033 & $-$0.0033 & $-$0.1436 & $-$0.1432 & $-$0.1436 & $-$0.1428 \\ 
      &    & (0.0227) & (0.0227) & (0.0227) & (0.0227) & (0.0219) & (0.0220) & (0.0222) & (0.0222) \\ 
      \cline{3-10}
      & \multirow{2}{*}{RMOM} & 0.0183 & 0.0184 & 0.0182 & 0.0182 & 0.1436 & 0.1432 & 0.1436 & 0.1428 \\ 
      &    & (0.0137) & (0.0137) & (0.0138) & (0.0138) & (0.0219) & (0.0220) & (0.0222) & (0.0222) \\ 
      \cline{2-10}
      & Estimation & 36.2 & 41.2 & 92.1 & 95.8 & 19.4 & 29.8 & 63.5 & 108.1 \\ 
      & time (sec) & (16.2) & (18.2) & (18.2) & (20.8) & (8.8) & (22.1) & (27.1) & (31.3) \\ 
  \hline 
  \end{tabular}}
  \label{Table:predsupp_full}
\end{table}

\begin{table}[htbp]
	\centering 
	\caption{Prediction performances of MLE and TLR approximation estimates for different $nb$ values, where the MLOE, MMOM, and RMOM criteria are computed by Stein's method.  $\text{Bias}(\cdot)$ means the estimate of the parameter minus its true value, while the estimation time means the computational time of the corresponding estimation. The value of MLOE for all cases ($\nu=0.5$ and $\nu=1.0$) is multiplied by $10^6$. }  
	\resizebox{\textwidth}{!}{
		\begin{tabular}{|c|c|c|cccc|c|cccc|}
			\hline 
			\multirow{3}{*}{$h_{\text{eff}}$} & \multirow{3}{*}{Mean (sd)} & \multicolumn{5}{c|}{$\nu=0.5$} & \multicolumn{5}{c|}{$\nu=1.0$} \\
			\cline{3-12}
			&    & \multirow{2}{*}{MLE} & \multicolumn{4}{c|}{Tile size ($nb$)} & \multirow{2}{*}{MLE} & \multicolumn{4}{c|}{Tile size ($nb$)} \\
			\cline{4-7} 
			\cline{9-12} 
			&	&	& 400 & 450 & 600 & 900 &  & 400 & 450 & 600 & 900 \\
			\hline 
			\multirow{6}{*}{$0.2$} & \multirow{2}{*}{MLOE $(\times 10^6)$} & 3.3803 & 3.3673 & 3.3673 & 3.3674 & 3.3673 & 1.7289 & 1.7076 & 1.7074 & 1.7076 & 1.7039 \\ 
			&    & (6.1122) & (6.0769) & (6.0769) & (6.0770) & (6.0770) & (2.9000) & (2.8383) & (2.8382) & (2.8383) & (2.8400) \\ 
			\cline{3-12}
			& \multirow{2}{*}{MMOM} & 0.0017 & 0.0011 & 0.0011 & 0.0011 & 0.0011 & 0.0019 & $-$0.0016 & $-$0.0016 & $-$0.0016 & $-$0.0016 \\ 
			&    & (0.0232) & (0.0232) & (0.0232) & (0.0232) & (0.0232) & (0.0240) & (0.0240) & (0.0240) & (0.0240) & (0.0240) \\ 
			\cline{3-12}
			& \multirow{2}{*}{RMOM} & 0.0185 & 0.0185 & 0.0185 & 0.0185 & 0.0185 & 0.0189 & 0.0190 & 0.0190 & 0.0190 & 0.0190 \\ 
			&    & (0.0141) & (0.0140) & (0.0140) & (0.0140) & (0.0140) & (0.0147) & (0.0146) & (0.0146) & (0.0146) & (0.0146) \\ 
			\hline 
			\multirow{6}{*}{$0.4$} & \multirow{2}{*}{MLOE $(\times 10^6)$} & 0.9145 & 0.9055 & 0.9056 & 0.9056 & 0.9055 & 0.3845 & 0.3566 & 0.3566 & 0.3566 & 0.3566 \\ 
			&    & (1.9099) & (1.8897) & (1.8897) & (1.8897) & (1.8897) & (0.6155) & (0.5499) & (0.5498) & (0.5499) & (0.5499) \\ 
			\cline{3-12}
			& \multirow{2}{*}{MMOM} & 0.0018 & 0.0005 & 0.0005 & 0.0005 & 0.0005 & 0.0019 & $-$0.0107 & $-$0.0107 & $-$0.0107 & $-$0.0107 \\ 
			&    & (0.0228) & (0.0228) & (0.0228) & (0.0228) & (0.0228) & (0.0231) & (0.0230) & (0.0230) & (0.0230) & (0.0230) \\ 
			\cline{3-12}
			& \multirow{2}{*}{RMOM} & 0.0183 & 0.0182 & 0.0182 & 0.0182 & 0.0182 & 0.0185 & 0.0206 & 0.0206 & 0.0206 & 0.0206 \\ 
			&    & (0.0136) & (0.0135) & (0.0135) & (0.0135) & (0.0135) & (0.0137) & (0.0147) & (0.0147) & (0.0147) & (0.0147) \\ 
			\hline 
			\multirow{6}{*}{$0.8$} & \multirow{2}{*}{MLOE $(\times 10^6)$} & 0.2933 & 0.2876 & 0.2876 & 0.2876 & 0.2876 & 0.0977 & 0.0721 & 0.0719 & 0.0720 & 0.0722 \\ 
			&    & (0.8364) & (0.8221) & (0.8221) & (0.8221) & (0.8221) & (0.1723) & (0.1168) & (0.1167) & (0.1166) & (0.1167) \\ 
			\cline{3-12}
			& \multirow{2}{*}{MMOM} & 0.0016 & $-$0.0007 & $-$0.0007 & $-$0.0007 & $-$0.0007 & 0.0017 & $-$0.0438 & $-$0.0438 & $-$0.0438 & $-$0.0438 \\ 
			&    & (0.0227) & (0.0227) & (0.0227) & (0.0227) & (0.0227) & (0.0227) & (0.0225) & (0.0225) & (0.0225) & (0.0225) \\ 
			\cline{3-12}
			& \multirow{2}{*}{RMOM} & 0.0182 & 0.0182 & 0.0182 & 0.0182 & 0.0182 & 0.0182 & 0.0446 & 0.0446 & 0.0446 & 0.0446 \\ 
			&    & (0.0135) & (0.0135) & (0.0135) & (0.0135) & (0.0135) & (0.0135) & (0.0210) & (0.0210) & (0.0210) & (0.0210) \\ 
			\hline 
			\multirow{6}{*}{$1.6$} & \multirow{2}{*}{MLOE $(\times 10^6)$} & 0.1066 & 0.1029 & 0.1029 & 0.1029 & 0.1029 & 0.0267 & 0.0110 & 0.0109 & 0.0109 & 0.0110 \\ 
			&    & (0.3810) & (0.3692) & (0.3692) & (0.3692) & (0.3692) & (0.0657) & (0.0236) & (0.0235) & (0.0235) & (0.0235) \\ 
			\cline{3-12}
			& \multirow{2}{*}{MMOM} & 0.0014 & $-$0.0033 & $-$0.0033 & $-$0.0033 & $-$0.0033 & 0.0014 & $-$0.1428 & $-$0.1428 & $-$0.1428 & $-$0.1428 \\ 
			&    & (0.0227) & (0.0227) & (0.0227) & (0.0227) & (0.0227) & (0.0227) & (0.0222) & (0.0223) & (0.0223) & (0.0223) \\ 
			\cline{3-12}
			& \multirow{2}{*}{RMOM} & 0.0181 & 0.0182 & 0.0182 & 0.0182 & 0.0182 & 0.0182 & 0.1428 & 0.1428 & 0.1428 & 0.1428 \\ 
			&    & (0.0136) & (0.0138) & (0.0138) & (0.0138) & (0.0138) & (0.0135) & (0.0222) & (0.0223) & (0.0223) & (0.0223) \\ 
			\hline 
	\end{tabular}
    }
	\label{Table:nb_estimate_full_stein} 
\end{table}

\begin{table}[htbp]
	\centering 
	\caption{Estimation and prediction performances of the exact MLE and TLR approximation estimates for different $tlr\_acc$ values, where the MLOE, MMOM, and RMOM criteria are computed by Stein's method.  $\text{Bias}(\cdot)$ means the estimate of the parameter minus its true value, while the estimation time means the computational time of the corresponding estimation. The value of MLOE for all cases ($\nu=0.5$ and $\nu=1.0$) is multiplied by $10^6$. The missing part in the table (-) means that the result is not available, because the covariance matrix is numerically non positive-definite. } 
	\resizebox{\textwidth}{!}{
		\begin{tabular}{|c|c|c|cccc|c|cccc|}
			\hline
			\multirow{3}{*}{$h_{\text{eff}}$} & \multirow{3}{*}{Mean (sd)} & \multicolumn{5}{c|}{$\nu=0.5$} & \multicolumn{5}{c|}{$\nu=1.0$} \\
			\cline{3-12}
			&    & \multirow{2}{*}{MLE} & \multicolumn{4}{c|}{TLR accuracy ($tlr\_acc$)} & \multirow{2}{*}{MLE} & \multicolumn{4}{c|}{TLR accuracy ($tlr\_acc$)} \\
			\cline{4-7} 
			\cline{9-12} 
			&	&	& $10^{-5}$ & $10^{-7}$ & $10^{-9}$ & $10^{-11}$ &  & $10^{-5}$ & $10^{-7}$ & $10^{-9}$ & $10^{-11}$ \\
			\hline
			\multirow{6}{*}{$0.2$} & \multirow{2}{*}{MLOE $(\times 10^6)$} & 3.3803 & 3.8010 & 3.3673 & 3.3673 & 3.3673 & 1.7289 & 2.0364 & 1.7077 & 1.7074 & 1.7076 \\ 
			&    & (6.1122) & (7.9488) & (6.0770) & (6.0769) & (6.0769) & (2.9000) & (2.9816) & (2.8388) & (2.8382) & (2.8383) \\ 
			\cline{3-12}
			& \multirow{2}{*}{MMOM} & 0.0017 & 0.0009 & 0.0011 & 0.0011 & 0.0011 & 0.0019 & $-$0.0022 & $-$0.0016 & $-$0.0016 & $-$0.0016 \\ 
			&    & (0.0232) & (0.0233) & (0.0232) & (0.0232) & (0.0232) & (0.0240) & (0.0243) & (0.0240) & (0.0240) & (0.0240) \\ 
			\cline{3-12}
			& \multirow{2}{*}{RMOM} & 0.0185 & 0.0186 & 0.0185 & 0.0185 & 0.0185 & 0.0189 & 0.0194 & 0.0190 & 0.0190 & 0.0190 \\ 
			&    & (0.0141) & (0.0140) & (0.0140) & (0.0140) & (0.0140) & (0.0147) & (0.0146) & (0.0146) & (0.0146) & (0.0146) \\ 
			\hline
			\multirow{6}{*}{$0.4$} & \multirow{2}{*}{MLOE $(\times 10^6)$} & 0.9145 & 0.8590 & 0.9053 & 0.9056 & 0.9055 & 0.3845 & - & 0.3578 & 0.3566 & 0.3566 \\ 
			&    & (1.9099) & (1.924) & (1.8876) & (1.8897) & (1.8897) & (0.6155) & - & (0.5485) & (0.5498) & (0.5498) \\ 
			\cline{3-12}
			& \multirow{2}{*}{MMOM} & 0.0018 & 0.0004 & 0.0005 & 0.0005 & 0.0005 & 0.0019 & - & $-$0.0107 & $-$0.0107 & $-$0.0107 \\ 
			&    & (0.0228) & (0.0227) & (0.0228) & (0.0228) & (0.0228) & (0.0231) & - & (0.0230) & (0.0230) & (0.0230) \\ 
			\cline{3-12}
			& \multirow{2}{*}{RMOM} & 0.0183 & 0.0182 & 0.0182 & 0.0182 & 0.0182 & 0.0185 & - & 0.0206 & 0.0206 & 0.0206 \\ 
			&    & (0.0136) & (0.0135) & (0.0135) & (0.0135) & (0.0135) & (0.0137) & - & (0.0147) & (0.0147) & (0.0147) \\ 
			\hline
			\multirow{6}{*}{$0.8$} & \multirow{2}{*}{MLOE $(\times 10^6)$} & 0.2933 & 0.1383 & 0.2869 & 0.2876 & 0.2876 & 0.0977 & - & 0.0560 & 0.0719 & 0.0721 \\ 
			&    & (0.8364) & (0.2608) & (0.8222) & (0.8221) & (0.8221) & (0.1723) & - & (0.1076) & (0.1167) & (0.1167) \\ 
			\cline{3-12}
			& \multirow{2}{*}{MMOM} & 0.0016 & $-$0.0028 & $-$0.0007 & $-$0.0007 & $-$0.0007 & 0.0017 & - & $-$0.0441 & $-$0.0438 & $-$0.0438 \\ 
			&    & (0.0227) & (0.0230) & (0.0227) & (0.0227) & (0.0227) & (0.0227) & - & (0.0227) & (0.0225) & (0.0225) \\ 
			\cline{3-12}
			& \multirow{2}{*}{RMOM} & 0.0182 & 0.0185 & 0.0182 & 0.0182 & 0.0182 & 0.0182 & - & 0.0448 & 0.0446 & 0.0446 \\ 
			&    & (0.0135) & (0.0139) & (0.0135) & (0.0135) & (0.0135) & (0.0135) & - & (0.0211) & (0.0210) & (0.0210) \\ 
			\hline
			\multirow{6}{*}{$1.6$} & \multirow{2}{*}{MLOE $(\times 10^6)$} & 0.1066 & 0.0059 & 0.1026 & 0.1029 & 0.1029 & 0.0267 & - & 0.0077 & 0.0110 & 0.0110 \\ 
			&    & (0.3810) & (0.0356) & (0.3693) & (0.3692) & (0.3692) & (0.0657) & - & (0.0142) & (0.0236) & (0.0236) \\ 
			\cline{3-12}
			& \multirow{2}{*}{MMOM} & 0.0014 & $-$0.0044 & $-$0.0033 & $-$0.0033 & $-$0.0033 & 0.0014 & - & $-$0.1436 & $-$0.1428 & $-$0.1428 \\ 
			&    & (0.0227) & (0.0227) & (0.0227) & (0.0227) & (0.0227) & (0.0227) & - & (0.0222) & (0.0222) & (0.0222) \\ 
			\cline{3-12}
			& \multirow{2}{*}{RMOM} & 0.0181 & 0.0183 & 0.0182 & 0.0182 & 0.0182 & 0.0182 & - & 0.1436 & 0.1428 & 0.1428 \\ 
			&    & (0.0136) & (0.0139) & (0.0138) & (0.0138) & (0.0138) & (0.0135) & - & (0.0222) & (0.0222) & (0.0222) \\ 
			\hline
	\end{tabular}}
	\label{Table:tlracc_estimate_full_stein} 
\end{table}

\begin{table}[htbp]
	\centering 
	\caption{Estimation and prediction performances of the exact MLE and TLR approximation estimates for different $opt\_tol$ values, where the MLOE, MMOM, and RMOM criteria are computed by Stein's method.  $\text{Bias}(\cdot)$ means the estimate of the parameter minus its true value, while the estimation time means the computational time of the corresponding estimation. The value of MLOE for all cases ($\nu=0.5$ and $\nu=1.0$) is multiplied by $10^6$. } 
	\resizebox{\textwidth}{!}{
		\begin{tabular}{|c|c|c|cccc|c|cccc|}
			\hline
			\multirow{3}{*}{$h_{\text{eff}}$} & \multirow{3}{*}{Mean (sd)} & \multicolumn{5}{c|}{$\nu=0.5$} & \multicolumn{5}{c|}{$\nu=1.0$} \\
			\cline{3-12}
			&    & \multirow{2}{*}{MLE} & \multicolumn{4}{c|}{Optimization tolerance ($opt\_tol$)} & \multirow{2}{*}{MLE} & \multicolumn{4}{c|}{Optimization tolerance ($opt\_tol$)} \\
			\cline{4-7} 
			\cline{9-12}
			&	&	& $10^{-3}$ & $10^{-6}$ & $10^{-9}$ & $10^{-12}$ &	& $10^{-3}$ & $10^{-6}$ & $10^{-9}$ & $10^{-12}$ \\
			\hline
			\multirow{6}{*}{$0.2$} & \multirow{2}{*}{MLOE $(\times 10^6)$} & 3.3803 & 19.1697 & 3.3673 & 3.3673 & 3.3673 & 1.7289 & 7.5960 & 1.7074 & 1.7076 & 1.7076 \\ 
			&    & (6.1122) & (16.0331) & (6.0769) & (6.0770) & (6.0770) & (2.9000) & (5.8122) & (2.8382) & (2.8383) & (2.8383) \\ 
			\cline{3-12}
			& \multirow{2}{*}{MMOM} & 0.0017 & $-$0.0061 & 0.0011 & 0.0011 & 0.0011 & 0.0019 & $-$0.0103 & $-$0.0016 & $-$0.0016 & $-$0.0016 \\ 
			&    & (0.0232) & (0.0234) & (0.0232) & (0.0232) & (0.0232) & (0.0240) & (0.0243) & (0.0240) & (0.0240) & (0.0240) \\ 
			\cline{3-12}
			& \multirow{2}{*}{RMOM} & 0.0185 & 0.0196 & 0.0185 & 0.0185 & 0.0185 & 0.0189 & 0.0215 & 0.0190 & 0.0190 & 0.0190 \\ 
			&    & (0.0141) & (0.0140) & (0.0140) & (0.0140) & (0.0140) & (0.0147) & (0.0153) & (0.0146) & (0.0146) & (0.0146) \\ 
			\hline 
			\multirow{6}{*}{$0.4$} & \multirow{2}{*}{MLOE $(\times 10^6)$} & 0.9145 & 0.3160 & 0.9056 & 0.9055 & 0.9055 & 0.3845 & 0.1556 & 0.3566 & 0.3566 & 0.3566 \\ 
			&    & (1.9099) & (1.0943) & (1.8897) & (1.8897) & (1.8897) & (0.6155) & (0.2432) & (0.5498) & (0.5498) & (0.5498) \\ 
			\cline{3-12}
			& \multirow{2}{*}{MMOM} & 0.0018 & $-$0.0009 & 0.0005 & 0.0005 & 0.0005 & 0.0019 & $-$0.0105 & $-$0.0107 & $-$0.0107 & $-$0.0107 \\ 
			&    & (0.0228) & (0.0225) & (0.0228) & (0.0228) & (0.0228) & (0.0231) & (0.0222) & (0.0230) & (0.0230) & (0.0230) \\ 
			\cline{3-12}
			& \multirow{2}{*}{RMOM} & 0.0183 & 0.0179 & 0.0182 & 0.0182 & 0.0182 & 0.0185 & 0.0198 & 0.0206 & 0.0206 & 0.0206 \\ 
			&    & (0.0136) & (0.0135) & (0.0135) & (0.0135) & (0.0135) & (0.0137) & (0.0145) & (0.0147) & (0.0147) & (0.0147) \\ 
			\hline 
			\multirow{6}{*}{$0.8$} & \multirow{2}{*}{MLOE $(\times 10^6)$} & 0.2933 & 0.2085 & 0.2876 & 0.2876 & 0.2876 & 0.0977 & 0.0106 & 0.0719 & 0.0720 & 0.0720 \\ 
			&    & (0.8364) & (0.5511) & (0.8221) & (0.8221) & (0.8221) & (0.1723) & (0.0086) & (0.1167) & (0.1167) & (0.1167) \\ 
			\cline{3-12}
			& \multirow{2}{*}{MMOM} & 0.0016 & $-$0.0018 & $-$0.0007 & $-$0.0007 & $-$0.0007 & 0.0017 & $-$0.0448 & $-$0.0438 & $-$0.0438 & $-$0.0438 \\ 
			&    & (0.0227) & (0.0229) & (0.0227) & (0.0227) & (0.0227) & (0.0227) & (0.0226) & (0.0225) & (0.0225) & (0.0225) \\ 
			\cline{3-12}
			& \multirow{2}{*}{RMOM} & 0.0182 & 0.0182 & 0.0182 & 0.0182 & 0.0182 & 0.0182 & 0.0455 & 0.0446 & 0.0446 & 0.0446 \\ 
			&    & (0.0135) & (0.0140) & (0.0135) & (0.0135) & (0.0135) & (0.0135) & (0.0210) & (0.0210) & (0.0210) & (0.0210) \\ 
			\hline 
			\multirow{6}{*}{$1.6$} & \multirow{2}{*}{MLOE $(\times 10^6)$} & 0.1066 & 0.0620 & 0.1029 & 0.1029 & 0.1029 & 0.0267 & 0.0117 & 0.0110 & 0.0110 & 0.0110 \\ 
			&    & (0.3810) & (0.3190) & (0.3692) & (0.3692) & (0.3692) & (0.0657) & (0.0446) & (0.0236) & (0.0236) & (0.0236) \\ 
			\cline{3-12}
			& \multirow{2}{*}{MMOM} & 0.0014 & $-$0.0038 & $-$0.0033 & $-$0.0033 & $-$0.0033 & 0.0014 & $-$0.1432 & $-$0.1428 & $-$0.1428 & $-$0.1428 \\ 
			&    & (0.0227) & (0.0227) & (0.0227) & (0.0227) & (0.0227) & (0.0227) & (0.0220) & (0.0222) & (0.0222) & (0.0222) \\ 
			\cline{3-12}
			& \multirow{2}{*}{RMOM} & 0.0181 & 0.0184 & 0.0182 & 0.0182 & 0.0182 & 0.0182 & 0.1432 & 0.1428 & 0.1428 & 0.1428 \\ 
			&    & (0.0136) & (0.0137) & (0.0138) & (0.0138) & (0.0138) & (0.0135) & (0.0220) & (0.0222) & (0.0222) & (0.0222) \\ 
			\hline 
	\end{tabular}}
	\label{Table:prectlr_estimate_full_stein} 
\end{table}

\begin{table}[htbp]
	\centering
	\caption{Prediction performance and the computational time for Tile Low-Rank approximations with different combinations of $tlr\_acc$ and $opt\_tol$, where $\nu$ is the smoothness parameter and $h_{\text{eff}}$ is the effective range. The MLOE, MMOM, and RMOM criteria are computed by Stein's method. The estimation time means the computational time of the corresponding estimation, while value of MLOE for all cases ($\nu=0.5$ and $\nu=1.0$) is multiplied by $10^6$. } 
	\resizebox{\textwidth}{!}{
		\begin{tabular}{|c|c|cccc|cccc|}
			\hline 
			\multirow{2}{*}{$h_{\text{eff}}$} & \multirow{2}{*}{Mean (sd)} & \multicolumn{4}{c|}{$(tlr\_acc, opt\_tol)$, $\nu=0.5$} & \multicolumn{4}{c|}{$(tlr\_acc, opt\_tol)$, $\nu=1.0$} \\
			\cline{3-10}
			&    & $(10^{-7}, 10^{-3})$ & $(10^{-9}, 10^{-3})$ & $(10^{-7}, 10^{-6})$ & $(10^{-9}, 10^{-6})$ & $(10^{-7}, 10^{-3})$ & $(10^{-9}, 10^{-3})$ & $(10^{-7}, 10^{-6})$ & $(10^{-9}, 10^{-6})$ \\
			\hline 
			\multirow{6}{*}{$0.2$} & \multirow{2}{*}{MLOE $(\times 10^6)$} & 18.7630 & 19.1697 & 3.3673 & 3.3673 & 7.6887 & 7.5960 & 1.7077 & 1.7074 \\ 
			&    & (15.8851) & (16.0331) & (6.0770) & (6.0769) & (5.7695) & (5.8122) & (2.8388) & (2.8382) \\ 
			\cline{3-10}
			& \multirow{2}{*}{MMOM} & $-$0.0062 & $-$0.0061 & 0.0011 & 0.0011 & $-$0.0106 & $-$0.0103 & $-$0.0016 & $-$0.0016 \\ 
			&    & (0.0232) & (0.0234) & (0.0232) & (0.0232) & (0.0246) & (0.0243) & (0.0240) & (0.0240) \\ 
			\cline{3-10}
			& \multirow{2}{*}{RMOM} & 0.0193 & 0.0196 & 0.0185 & 0.0185 & 0.0218 & 0.0215 & 0.0190 & 0.0190 \\ 
			&    & (0.0142) & (0.0140) & (0.0140) & (0.0140) & (0.0154) & (0.0153) & (0.0146) & (0.0146) \\ 
			\hline 
			\multirow{6}{*}{$0.4$} & \multirow{2}{*}{MLOE $(\times 10^6)$} & 0.3203 & 0.3160 & 0.9053 & 0.9056 & 0.1547 & 0.1556 & 0.3578 & 0.3566 \\ 
			&    & (1.1115) & (1.0943) & (1.8876) & (1.8897) & (0.2434) & (0.2432) & (0.5485) & (0.5498) \\ 
			\cline{3-10}
			& \multirow{2}{*}{MMOM} & $-$0.0011 & $-$0.0009 & 0.0005 & 0.0005 & $-$0.0105 & $-$0.0105 & $-$0.0107 & $-$0.0107 \\ 
			&    & (0.0226) & (0.0225) & (0.0228) & (0.0228) & (0.0221) & (0.0222) & (0.0230) & (0.0230) \\ 
			\cline{3-10}
			& \multirow{2}{*}{RMOM} & 0.0180 & 0.0179 & 0.0182 & 0.0182 & 0.0198 & 0.0198 & 0.0206 & 0.0206 \\ 
			&    & (0.0136) & (0.0135) & (0.0135) & (0.0135) & (0.0143) & (0.0145) & (0.0147) & (0.0147) \\ 
			\hline 
			\multirow{6}{*}{$0.8$} & \multirow{2}{*}{MLOE $(\times 10^6)$} & 0.1643 & 0.2085 & 0.2869 & 0.2876 & 0.0109 & 0.0106 & 0.0560 & 0.0719 \\ 
			&    & (0.3569) & (0.5511) & (0.8222) & (0.8221) & (0.0093) & (0.0086) & (0.1076) & (0.1167) \\ 
			\cline{3-10}
			& \multirow{2}{*}{MMOM} & $-$0.0019 & $-$0.0018 & $-$0.0007 & $-$0.0007 & $-$0.0447 & $-$0.0448 & $-$0.0441 & $-$0.0438 \\ 
			&    & (0.0230) & (0.0229) & (0.0227) & (0.0227) & (0.0227) & (0.0226) & (0.0227) & (0.0225) \\ 
			\cline{3-10}
			& \multirow{2}{*}{RMOM} & 0.0183 & 0.0182 & 0.0182 & 0.0182 & 0.0455 & 0.0455 & 0.0448 & 0.0446 \\ 
			&    & (0.0140) & (0.0140) & (0.0135) & (0.0135) & (0.0211) & (0.0210) & (0.0211) & (0.0210) \\ 
			\hline 
			\multirow{6}{*}{$1.6$} & \multirow{2}{*}{MLOE $(\times 10^6)$} & 0.0649 & 0.0620 & 0.1026 & 0.1029 & 0.0111 & 0.0117 & 0.0077 & 0.0110 \\ 
			&    & (0.2983) & (0.3190) & (0.3693) & (0.3692) & (0.0424) & (0.0446) & (0.0142) & (0.0236) \\ 
			\cline{3-10}
			& \multirow{2}{*}{MMOM} & $-$0.0038 & $-$0.0038 & $-$0.0033 & $-$0.0033 & $-$0.1436 & $-$0.1432 & $-$0.1436 & $-$0.1428 \\ 
			&    & (0.0227) & (0.0227) & (0.0227) & (0.0227) & (0.0219) & (0.0220) & (0.0222) & (0.0222) \\ 
			\cline{3-10}
			& \multirow{2}{*}{RMOM} & 0.0183 & 0.0184 & 0.0182 & 0.0182 & 0.1436 & 0.1432 & 0.1436 & 0.1428 \\ 
			&    & (0.0137) & (0.0137) & (0.0138) & (0.0138) & (0.0219) & (0.0220) & (0.0222) & (0.0222) \\ 
			\hline 
	\end{tabular}}
	\label{Table:predsupp_full_stein}
\end{table}

\end{document}